\documentclass[12pt,draftcls,journal,letterpaper,onecolumn,twoside]{IEEEtran}  
\usepackage{graphicx}
\usepackage[cmex10]{amsmath}
\usepackage{amsfonts}
\usepackage{amssymb}
\usepackage{mathrsfs}
\usepackage{epstopdf}
\usepackage{color}
\usepackage{cite}
\newcommand{\rank}{{\mbox{\textrm{Rank}}}}

\newcommand{\bW}{\mathbf{W}}

\newcommand{\bw}{\mathbf{w}}

\begin{document}
\title{Outage Constrained Robust Secure Transmission for  MISO Wiretap Channels }
%
%
%
\author{Shuai~Ma,  Mingyi Hong, Enbin Song, Xiangfeng Wang and ~Dechun~Sun
\thanks{S. Ma and D. Sun  are with the State Key Laboratory of Integrated Services
Networks (ISN Lab), Xidian University, Xi'an, 710071,
China (e-mail: mashuai@stu.xidian.edu.cn; dechsun@sina.com).}
\thanks{ M. Hong (Corresponding author)
is with the Department of Electrical and
Computer Engineering University of Minnesota, Minneapolis, MN 55455,
USA (e-mails: mhong@umn.edu).}
\thanks{E. Song  is with the Department of Mathematics,
Sichuan University, Chengdu 610064, China (e-mail: e.b.song@163.com).}
\thanks{X. Wang is with the Department of Mathematics, Nanjing University, 22 Hankou Road, Nanjing, 210093,   China (e-mail: xfwang.nju@gmail.com).}
}

\maketitle
\begin{abstract}

In this paper we consider the robust secure beamformer design for MISO wiretap channels. Assume that the eavesdroppers' channels are only partially available at the transmitter, we seek to maximize the secrecy rate under the transmit power and secrecy rate outage probability constraint. The outage probability constraint requires that the secrecy rate  exceeds  certain threshold with high probability. Therefore including such constraint in the design naturally ensures the desired robustness. Unfortunately, the presence of the probabilistic constraints makes the problem non-convex and hence difficult to solve. In this paper, we investigate the outage probability constrained secrecy rate maximization problem using a novel two-step approach. Under a wide range of uncertainty models, our developed algorithms can obtain high-quality solutions, sometimes even exact global solutions, for the robust secure beamformer design problem.
Simulation results are presented to verify the effectiveness and robustness of the proposed algorithms.

\end{abstract}
\begin{IEEEkeywords}
Physical-layer secrecy, MISO wiretap channel,  Robust secrecy beamforming,  Chance constraints.
\end{IEEEkeywords}

\IEEEpeerreviewmaketitle

\section{Introduction}

Wireless communication is susceptible to eavesdropping due to its broadcast nature.
Traditionally, security  is treated in cryptography through data-encryption at the application
layer. However, the open nature of wireless medium and the dynamic topology of mobile networks may introduce significant challenges to secret key transmission and management \cite{Debbah,Liang}.
In comparison to the conventional cryptographic approaches, physical-layer secrecy can achieve perfect security without using an encryption key. The information-theoretic notion of security is introduced by
Shannon to study secure communication over point-to-point noiseless channels \cite{Shannon}.
Wyner defined the secrecy capacity for a wiretap channels as the
upper bound of all achievable rates in which private messages are guaranteed to be decoded by the legitimate receiver, while being kept perfectly secret from the eavesdropper \cite{Wyner}.

In a wiretap channel, to guarantee non-zero secrecy rate, the eavesdropper's channel should be worse than the legitimate's channel \cite{Wyner}. However, this may not always be possible in practical wireless
environment. By utilizing multiple antennas at the transmitter, the dependence on channel conditions can be greatly reduced. This can be attribute to the extra spatial degrees of freedom provided by the antennas arrays, which enables the transmitter to further degrade the reception of the eavesdroppers while at the same time enhance the rate of the desired receiver. Recently, considerable research has investigated optimization algorithms for improving secrecy rate in wiretap channels with multiple antennas \cite{Liu,Khisti,Khisti2,Trappe,Shafiee,Li,Hong}.

There are roughly two approaches for designing transmission schemes in the presence of multiple transmit antennas: 1) single-stream transmit beamforming, in which the transmit signal is steered towards the legitimate receiver, while the power leakage to the eavesdroppers is reduced at the same time; 2) joint beamforming and artificial noise (AN) generation, in which the transmit power is split into a data stream and an   AN \cite{Negi,Fakoorian,Liao,Li4}. The AN is used to generate interference to degrade the reception quality at the eavesdropper. In this paper, we focus on the single-stream transmit beamforming approach.
The secrecy capacity of the multiple-input single-output (MISO) wiretap channel was proved in  \cite{Shafiee}.
The authors in \cite{Khisti} investigated the fading MISO wiretap channel, and the analysis was extended to the multi-input multi-output (MIMO) case in \cite{Khisti2}.
The secrecy capacity for a Gaussian broadcast channel was computed in \cite{Liu}, where a multi-antenna transmitter sends independent confidential messages to two users. We note that all the above results are based on the somewhat unrealistic assumption that the channel state information (CSI) of {\it both} legitimate receiver and the eavesdropper is perfect known to the transmitter. However, in practice perfect CSI of the legitimate user is already sometimes difficult to obtain (due to estimation errors or feedback errors), not to mention that of the eavesdroppers.
Naturally, such CSI uncertainty   heavily  deteriorates  the performance of the system \cite{Shi}.

 Motivated by this fact, the robust design for physical-layer secrecy with imperfect CSI has received a lot of attention recently. In \cite{Shi,Wolf,Zhang}, the problem of maximizing the worst-case secrecy rate under various scenarios was studied,  with imperfect eavesdroppers' CSI (ECSI) and perfect legitimate receivers' CSI (LCSI). Under the assumption of norm-bounded uncertainty, the  secrecy rate maximization problem with both imperfect  ECSI and imperfect  LCSI was investigated in   \cite{Li3,Huang}.
 It is worth noting that all the above mentioned works focus on bounded CSI errors using the worst case approach.  Although such approach guarantees the performance of the worst CSI errors scenarios, it often leads to a very conservative design, because the extreme conditions may rarely occur.  On the other hand, the robustness of the design can also be improved by introducing certain outage probability constraints, which often yields less conservative results.   A detailed characterization of the outage secrecy capacity of slow fading single-input single-output (SISO) wiretap channels was provided in  \cite{Barros}, where only the LCSI is known exactly.
 In \cite{Parada}, the authors  investigated  a single letter characterization of the
secrecy capacity of the single-input multiple-outputs (SIMO) channel and the impact of slow fading on the secrecy capacity. With imperfect ECSI and perfect LCSI,  the authors in \cite{Jorswieck} proposed to minimize the outage probability of secure transmission for both cases of single-stream transmit beamforming  and AN aided transmit beamforming.

In this paper, we seek to design robust secure beamforming strategies for MISO wiretap channels under various assumptions on the CSI.
In particular, we consider three CSI uncertainty scenarios: (a) perfect LCSI and statistical ECSI; (b) perfect LCSI and imperfect ECSI, and (c) imperfect LCSI and imperfect ECSI. Here {\it imperfect CIS} refers to the case where the channel lies in some uncertainty set centered at the true channel; {\it statistical CSI} means only the distribution of the channel is available. In each of the considered cases, the presence of channel uncertainty leads to the outage event. That is, any given secrecy  rate  requirement  cannot be  guaranteed  all the time. 
Therefore, we focus on studying the secrecy rate maximization problem with a given secrecy outage probability. In other words, we design robust secure  beamformer in a way that ensures the probability that an outage event occurs is smaller than certain given threshold. Unfortunately, in general the probabilistic constraints often have no closed-form expressions and are seldom convex \cite{Ben,Stochastic}.

The main contribution of this paper is the development of a suite of algorithms that handle the difficult outage probability constraint for all three CSI uncertainty scenarios. Our first step is to decompose the problem into a sequence of power minimization problems under the secrecy outage constraints. Then we propose three new algorithms to solve the resulting outage probability constrained power minimization problem, one for each  scenario:
\begin{enumerate}

\item {\bf Perfect LCSI and statistical ECSI (Scenario 1)}: The chance constrained power minimization problem is first equivalently converted into a deterministic problem.
For the case with a single eavesdropper, we derived the optimal solution in closed form, while in the presence of multiple eavesdroppers, the problem is solved by using semidefinite relaxation (SDR). Importantly, we show that in the latter case, whenever the original problem is feasible, the SDR is always tight.


\item {\bf Perfect LCSI and imperfect ECSI (Scenario 2)}: The chance constrained power minimization problem is first lifted into high dimensions, and then conservatively transformed into a convex SDP by using the the Bernstein-type Inequality I \cite{Bechar,Ma2}. A customized procedure: Projection Approximation Procedure  is then developed to recover a high quality rank-1 solution of the original problem.


 \item {\bf Imperfect LCSI and imperfect ECSI (Scenario 3)}: In this case, there are multiple types of CSI uncertainties in the chance constraint.  We first recombine the CSI errors to higher dimension.  We then conservatively transform the power minimization problem into a deterministic form by using the the Bernstein-type Inequality II  \cite{Bechar,Wang_a,Wang_b,Ma2}, for which the SDR is used again to relax the deterministic problem into a convex SDP problem.

\end{enumerate}


\emph{Notations}: Boldfaced lowercase (resp. uppercase) letters are
used to represent vectors (resp. matrices). All vectors are column vectors.  The symbols ${\left( {\cdot} \right)^*}$, ${\left(
{\cdot} \right)^T}$, ${\left(
{\cdot} \right)^H}$, ${\mathbb{C}^N}$, ${\text{Tr}}\left(  \cdot  \right)$, $\left\| {\cdot} \right\|$, $\odot$ and $\otimes$
    denote respectively conjugate, transpose, conjugate transpose, the space of $N \times 1$
complex vector, the trace,  the Frobenius norm, the Hadamard product and Kronecker  product. $\operatorname{Re} \left\{ {\cdot} \right\}$ extracts
the real part of its argument.  ${\mathbf{{x}}} \sim {\mathcal{CN}}\left(
{\mathbf{{m}},\mathbf{{V}}} \right)$ means that $\mathbf{{x}}$ is
complex Gaussian distributed with mean vector $\mathbf{{m}}$ and
covariance matrix $\mathbf{{V}}$. $\rho \left( {{{\mathbf{A}}}} \right)$ denotes the largest eigenvalue of the matrix ${{{\mathbf{A}}}}$.



 \section{System Model and Problem Formulation}
\subsection{System Model}
\begin{figure}[htbp]
\centering
\includegraphics[width=5.5cm]{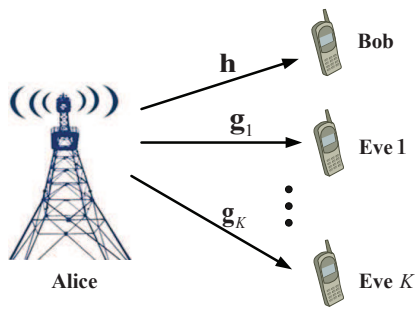}
\caption{System model.} \label{fig_sim}
\end{figure}
We consider a MISO communication system with a source
node (Alice),  a destination node (Bob), and multiple
eavesdroppers (Eves), as shown in Fig. 1. Suppose that Alice has  ${N_t}$ transmit antennas,
while both Bob and Eves have a single receive antenna.
In this model, Alice sends private messages to Bob in the presence
of Eves, who are able to eavesdrop on the link between Alice
and Bob.  Assuming that channels are flat-fading, the signals
received by Bob and Eves are given by
\begin{subequations}
\begin{align}
  &{{\mathbf{y}}_b}\left( t \right) = {{\mathbf{h}}^H}{\mathbf{w}}s\left( t \right) + {n_b}\left( t \right), \label{eq:1a} \\
  &{{\mathbf{y}}_{e,k}}\left( t \right) = {\mathbf{g}}_k^H{\mathbf{w}}s\left( t \right) + {n_k}\left( t \right),\forall k \in \mathcal{K}, \label{eq:1b}
\end{align}
\end{subequations}
where $s\left( t \right)$
is the data stream intended for Bob, with ${\rm E}\left\{ {{{\left| {s\left( t \right)} \right|}^2}} \right\} = 1$; ${\mathbf{w}} \in {\mathbb{C}^{{N_t}}}$ is the transmit beamformer vector for ${s\left( t \right)}$;
  ${\mathbf{h}} \in {\mathbb{C}^{{N_t}}}$
 is the channel from Alice to Bob, ${{\mathbf{g}}_k} \in {\mathbb{C}^{{N_t}}}$
is the channel from Alice to the $k$th Eve;
$n_b\left( t \right)$ and ${n_k}\left( t \right)$
are independent identically distributed (i.i.d.) circularly
symmetric complex-valued Gaussian noises:
$n_b\left( t \right) \sim \mathcal{CN}\left( {0,{\delta_b ^2}} \right)$ and
${n_k}\left( t \right) \sim \mathcal{CN}\left( {0,\delta _{e,k}^2} \right)$;  $\mathcal{K} = \left\{ {1,2,...,K} \right\}$.
The received signal-to-noise ratio (SNR) at Bob is given by
\begin{align}
{\mathbf{SN}}{{\mathbf{R}}_b}\left( {\mathbf{w}} \right) = \frac{{{\text{E}}\left\{ {{{\left\| {{{\mathbf{h}}^H}{\mathbf{w}}s\left( t \right)} \right\|}^2}} \right\}}}
{{\delta _b^2}} = \frac{{{{\left\| {{{\mathbf{h}}^H}{\mathbf{w}}} \right\|}^2}}}
{{\delta _b^2}}.\label{eq:2}
\end{align}
Likewise, the received SNR  at the $k$th Eve can be expressed as
\begin{align}{\mathbf{SN}}{{\mathbf{R}}_{e,k}}\left( {\mathbf{w}} \right) = \frac{{{\text{E}}\left\{ {{{\left\| {{\mathbf{g}}_k^H{\mathbf{w}}s\left( t \right)} \right\|}^2}} \right\}}}
{{\delta _{e,k}^2}} = \frac{{{{\left\| {{\mathbf{g}}_k^H{\mathbf{w}}} \right\|}^2}}}
{{\delta _{e,k}^2}}.\label{eq:3}
\end{align}

The average transmit power of Alice   is 
\begin{align}{\text{E}}\left\{ {{{\left\| {{\mathbf{w}}s\left( t \right)} \right\|}^2}} \right\} = {\left\| {\mathbf{w}} \right\|^2}.\label{eq:4}\end{align}

According to \cite{Shafiee,Khisti}, the instantaneous secrecy rate
is \begin{align}{R} = {\left[ {{{\log }_2}\left( {1 + {\mathbf{SN}}{{\mathbf{R}}_b}\left( {\mathbf{w}} \right)} \right) - \mathop {\max }\limits_{k \in \mathcal{K}} {{\log }_2}\left( {1 + {\mathbf{SN}}{{\mathbf{R}}_{e,k}}\left( {\mathbf{w}} \right)} \right)} \right]^ + .}\label{eq:5}\end{align}

A commonly used criteria for designing the transmit strategy is to maximize the achievable secrecy rate, subject to a total power constraint \cite{Hong}
 \begin{subequations}\label{eq:original_problem_K}
\begin{align}
  \mathop {\max }\limits_{{\mathbf{w}},R}&\quad R \\
  {\text{s}}{\text{.t}}{\text{.}}& \quad {{{\log }_2}\left( {1 + \frac{{{{\left\| {{{\mathbf{h}}^H}{\mathbf{w}}} \right\|}^2}}}
{{\delta _b^2}}} \right) -  {{\log }_2}\left( {1 + \frac{{{{\left\| {{\mathbf{g}}_k^H{\mathbf{w}}} \right\|}^2}}}
{{\delta _{e,k}^2}}} \right)} { \geq  R},\ \forall~k\in\mathcal{K},   \\
  &{\left\| {\mathbf{w}} \right\|^2} \leq  P,
\end{align}
\end{subequations}
where $P$ is the given average transmit power limit for Alice.
\subsection{CSI Uncertainty Scenarios}\label{sub:CSI}
One important factor that affects the above secrecy rate maximization problem is the availability of CSI. In most cases, the CSI between Alice and the legitimate receiver Bob can be quite accurate, as it is usually learned at both the receiver side and the transmitter side by training and feedback. However, the CSI between Alice and  Eve is rarely so, due to the limited cooperation among them for estimating the channel. As a result, any practical design to achieve high secrecy rate must take CSI uncertainty into consideration. In this work, we consider the following three scenarios that cover a wide range of CSI uncertainties.

\subsubsection{ Scenario 1: Perfect LCSI  and statistical ECSI}

We first consider a scenario that often arises in practice, in which Eves are not part of the legitimate system, hence their channels are not known. That is, Alice  knows the full CSI of the channel $\mathbf{h}$ but only some statistical information about ECSI \cite{Shafiee,Li}:
 \begin{align}
&{{\mathbf{g}}_k} \sim \mathcal{CN}\left( {0,{{\mathbf{G}}_k}} \right),\forall k \in \mathcal{K},\label{eq:6}
\end{align}
where ${{\mathbf{G}}_k} \succ {\mathbf{0}}$.  Note that in \cite{Shafiee,Li}, similar scenarios are considered, but with the important difference that only the nonrobust ergodic secrecy rate maximization problem is investigated.

\subsubsection{ Scenario 2: Perfect LCSI and imperfect  ECSI}

 Consider the scenario where Eves are regular users of the
system, but the cooperation between Alice and Eves is limited so that Alice only has some imprecise knowledge about the channel to Eves:
   \begin{align}
&{{\mathbf{g}}_k} = {\widehat{\mathbf{g}}_k}+ \Delta {{\mathbf{g}}_k},\forall k \in \mathcal{K},\label{eq:7}
\end{align}
where  ${\widehat {{\mathbf{{g}}}}_{k}} \in
 {\mathbb{C}^{{N_t}}}$ is the estimated CSI,  $\Delta {{\mathbf{{g}}}_{k}}$ is the stochastic CSI errors, following the distribution $\Delta {{\mathbf{g}}_k} \sim \mathcal{C}\mathcal{N}\left( {{\mathbf{0}},{{\mathbf{E}}_{e,k}}} \right)$, with ${{\mathbf{E}}_{e,k}} \succ {\mathbf{0}}$. Such uncertainty model has been considered in  \cite{Jorswieck}, but with a different design objective (minimize the outage probability) and only a single Eve.

  \subsubsection{ Scenario 3: Imperfect LCSI and imperfect  ECSI}

 We consider the case in which Eves are parts of the communication system \cite{Liang,Trappe}.
 Differently from the previous case, we model the CSIs for {\it both} Bob and Eves as being imperfect \cite{Li3}:
  \begin{align}
{\mathbf{h}} = \widehat{\mathbf{h}}{\text{  +  }}\Delta {\mathbf{h}},\ {{\mathbf{g}}_k} = {\widehat{\mathbf{g}}_k}{\text{  +  }}\Delta {{\mathbf{g}}_k},\forall k \in \mathcal{K},\label{eq:8}
\end{align}
where   ${\widehat {{\mathbf{{h}}}}} \in
 {\mathbb{C}^{{N_t}}}$ and ${\widehat {{\mathbf{{g}}}}_{k}} \in
 {\mathbb{C}^{{N_t}}}$ are the estimated CSI;  $\Delta {{\mathbf{{h}}}}$ and $\Delta {{\mathbf{{g}}}_{k}}$ are the  corresponding stochastic CSI errors, which respectively follows the distribution
$\Delta {{\mathbf{h}}} \sim \mathcal{C}\mathcal{N}\left( {{\mathbf{0}},{{\mathbf{E}}_{{{{b}}}}}} \right)$, ${{\mathbf{E}}_{{{{b}}}}} \succ {\mathbf{0}}$, and $\Delta {{\mathbf{g}}_k} \sim \mathcal{C}\mathcal{N}\left( {{\mathbf{0}},{{\mathbf{E}}_{e,k}}} \right)$, ${{\mathbf{E}}_{e,k}} \succ {\mathbf{0}}$.

\textbf{Remark 1} (Choice of error models):
 We have used Gaussian random vectors to model the imperfect CSI in Scenario 2 and 3.
The reason that we choose such model as opposed to characterizing the error as bounded random variables (see, e.g., \cite{Palomar2}) is given below. In the process of acquiring the CSI by the Alice,  there are two main sources of CSI errors: the estimation error and the quantization error. We consider the case that the estimation is  not very accurate but the amount of bits available for feeding back the CSI (which determine the size of the quantization codebook) is sufficient. Therefore the estimation error is the dominant factor for the uncertainty of the CSI. It is  known that when estimating channels using the MMSE method, the  CSI  errors tend to follow Gaussian distribution. We mention that the above model has already been used in \cite{Palomar2,Chung,Du,Palomar,Ottersten} to model CSI errors arise in other communication systems.

\subsection{Problem Formulation} \label{sub:formulation}

In all the uncertainty models presented above, limited CSI knowledge makes it difficult to design a transmit strategy that is able to guarantee a given rate target ${{R}}>0$ all the time. Fortunately, in practice many wireless applications (such as video streaming, voice over IP) are able to tolerate occasional events of outage without significantly affecting users' QoS \cite{Jorswieck}. Therefore it is reasonable to design transmit strategies that can meet the users' rate requirement with a high probability. Formally, we are interested in solving the following chance constrained program (which is a modification of problem \eqref{eq:original_problem_K})
 \begin{subequations}\label{eq:chance_constrained}
\begin{align}
  \mathop {\max }\limits_{{\mathbf{w}},R}&\quad R \label{eq:9a} \\
  {\text{s}}{\text{.t}}{\text{.}}&  \Pr \Bigg{\{} {{{\log }_2}\left( {1 + \frac{{{{\left\| {{{\mathbf{h}}^H}{\mathbf{w}}} \right\|}^2}}}
{{\delta _b^2}}} \right) -  {{\log }_2}\left( {1 + \frac{{{{\left\| {{\mathbf{g}}_k^H{\mathbf{w}}} \right\|}^2}}}
{{\delta _{e,k}^2}}} \right)} { \geq  R} \Bigg{\}} \geq  1 - {p_{k,out}},~ \forall k \in \mathcal{K}, \label{eq:9b} \\
  &\quad{\left\| {\mathbf{w}} \right\|^2} \leq  P, \label{eq:9c}
\end{align}
\end{subequations}
where ${p_{k,out}} \in \left( {0,1} \right]$ is the maximum allowable secrecy outage probability for the $k$th Eve.

The  chance constrained robust  beamforming design \eqref{eq:chance_constrained} is non-convex, and thus is not likely to  be solved efficiently. To make the problem tractable,  we first decompose \eqref{eq:chance_constrained} into a sequence
of probability constrained power minimization problems, one for each target rate $R>0$:
\begin{subequations}\label{eq:power_min}
\begin{align}
  &\mathop {\min }\limits_{\mathbf{w}}~ {\left\| {\mathbf{w}} \right\|^2} \label{eq:11a} \\
  {\text{s}}{\text{.t}}{\text{.}}&  \Pr \Bigg{\{} {{{\log }_2}\left( {1 + \frac{{{{\left\| {{{\mathbf{h}}^H}{\mathbf{w}}} \right\|}^2}}}
{{\delta _b^2}}} \right) -  {{\log }_2}\left( {1 + \frac{{{{\left\| {{\mathbf{g}}_k^H{\mathbf{w}}} \right\|}^2}}}
{{\delta _{e,k}^2}}} \right)}  { \geq  R} \Bigg{\}} \geq  1 - {p_{k,out}},~ \forall k \in \mathcal{K}. \label{eq:11b}
\end{align}
\end{subequations}

Obviously, the  optimal  objective value of  the above problem is monotonically increasing with respect to ${{R}^{{\text{opt}}}} $. 
Thus, by solving the problem \eqref{eq:power_min} with different ${R}$ and using a bisection search \cite{Boyd} over ${R}$, ${{R}^{{\text{opt}}}}$ can be obtained. In the subsequent sections, we will focus on solving \eqref{eq:power_min} for different uncertainty models. 

\section{Proposed Methods}

\subsection {Scenario 1: Perfect LCSI  and statistical ECSI}

In this scenario, only the statistical ECSI of the form ${{\mathbf{g}}_k} \sim \mathcal{CN}\left( {0,{{\mathbf{G}}_k}} \right),\forall k \in \mathcal{K}$ is known to Alice. Therefore the left hand side of the constraint \eqref{eq:11b} can be reformulated as
\begin{subequations}
\begin{align}
  &\Pr \left\{ {{{\log }_2}\left( {1 + \frac{{{{\left\| {{{\mathbf{h}}^H}{\mathbf{w}}} \right\|}^2}}}
{{\delta _b^2}}} \right) - {{\log }_2}\left( {1 + \frac{{{{\left\| {{\mathbf{g}}_k^H{\mathbf{w}}} \right\|}^2}}}
{{\delta _{e,k}^2}}} \right) \geq R} \right\}\nonumber\\
  =&\Pr \Bigg{\{}{{{\log }_2}\left( {1 + \frac{{{{\mathbf{w}}^H}{\mathbf{h}}{{\mathbf{h}}^H}{\mathbf{w}}}}
{{\delta _b^2}}} \right) - {{\log }_2}\left( {1 + \frac{{{{\mathbf{w}}^H}{{\mathbf{g}}_k}{\mathbf{g}}_k^H{\mathbf{w}}}}
{{\delta _{e,k}^2}}} \right)} { \geq  R} \Bigg{\}}\label{eq:12a}\\
   = &\Pr \left\{ {\frac{{\delta _{e,k}^2\left( {\delta _b^2 + {{\mathbf{w}}^H}{\mathbf{h}}{{\mathbf{h}}^H}{\mathbf{w}}} \right)}}
{{\delta _b^2\left( {\delta _{e,k}^2 + {{\mathbf{w}}^H}{{\mathbf{g}}_k}{\mathbf{g}}_k^H{\mathbf{w}}} \right)}} \geq {2^R}} \right\} \label{eq:12b} \\
   = &\Pr \left\{ {{{\mathbf{w}}^H}{{\mathbf{g}}_k}{\mathbf{g}}_k^H{\mathbf{w}} \leq \delta _{e,k}^2\left( {\frac{{\delta _b^2 + {{\mathbf{w}}^H}{\mathbf{h}}{{\mathbf{h}}^H}{\mathbf{w}}}}
{{\delta _b^2{2^R}}} - 1} \right)} \right\}\label{eq:12c}\\
   =& 1 - \exp \left( {\frac{{\delta _{e,k}^2}}
{{{{\mathbf{w}}^H}{{\mathbf{G}}_{k}}{\mathbf{w}}}}\left( {1 - \frac{{\delta _b^2 + {{\mathbf{w}}^H}{\mathbf{h}}{{\mathbf{h}}^H}{\mathbf{w}}}}
{{\delta _b^2{2^R}}}} \right)} \right). \label{eq:12d}
\end{align}
\end{subequations}

The equality in \eqref{eq:12d} holds true due to  the fact that  the random variable  ${{\mathbf{w}}^H}{{\mathbf{g}}_k}{\mathbf{g}}_k^H{\mathbf{w}}$ follows exponential distribution with mean ${{{{\mathbf{w}}^H}{{\mathbf{G}}_{k}}{\mathbf{w}}}}$ \cite{Kandukuri}.

Substituting \eqref{eq:12d} into \eqref{eq:11b}, we have
\begin{align}1 - \exp \left( {\frac{{\delta _{e,k}^2}}
{{{{\mathbf{w}}^H}{{\mathbf{G}}_{k}}{\mathbf{w}}}}\left( {1 - \frac{{\delta _b^2 + {{\mathbf{w}}^H}{\mathbf{h}}{{\mathbf{h}}^H}{\mathbf{w}}}}
{{\delta _b^2{2^R}}}} \right)} \right) \geq  1 - {p_{k,out}},\label{eq:13}\end{align}
which is equivalent to
\begin{align}\delta _{e,k}^2\left( {1 - \frac{1}
{{{2^R}}}} \right) \leq  {{\mathbf{w}}^H}\left( {{{\mathbf{G}}_{k}}\ln {p_{k,out}} + \frac{{\delta _{e,k}^2}}
{{\delta _b^2{2^R}}}{\mathbf{h}}{{\mathbf{h}}^H}} \right){\mathbf{w}}.\label{eq:14}
\end{align}

  We conclude that for scenario 1, the  problem \eqref{eq:power_min} is equivalent to the following deterministic problem:
  \begin{subequations}\label{eq:scenario1}
\begin{align}
  \mathop {\min }\limits_{\mathbf{w}}&\quad {\left\| {\mathbf{w}} \right\|^2} \label{eq:15a}\\
  {\text{s}}{\text{.t}}{\text{.}}&\quad{{\mathbf{w}}^H}\left( {{{\mathbf{G}}_{k}}\ln {p_{k,out}} + \frac{{\delta _{e,k}^2}}
{{\delta _b^2{2^R}}}{\mathbf{h}}{{\mathbf{h}}^H}} \right){\mathbf{w}} \geq  \delta _{e,k}^2\left( {1 - \frac{1}
{{{2^R}}}} \right), \forall k \in \mathcal{K}, \label{eq:15b}
\end{align}
\end{subequations}
The above problem is a nonconvex quadratically constrained quadratic problem (QCQP), where the nonconvexity comes from the (possibly indefinite) quadratic constraints \eqref{eq:15b}. The following series of results characterize its feasibility conditions.


%

\textbf{Proposition 1:}
{\it For scenario 1, when there is a single eavesdropper (i.e., $K=1$), the necessary and sufficient condition for problem \eqref{eq:scenario1} to be feasible is $\rho \left( {\Lambda} \right) > 0$,
 where $\Lambda  \triangleq {{\mathbf{G}}_1}\ln {p_{k,out}} + \frac{{\delta _{e,1}^2}}
{{\delta _b^2{2^R}}}{\mathbf{h}}{{\mathbf{h}}^H}$.
When condition $\rho \left( {\Lambda} \right) > 0$ is satisfied, the optimal solution to \eqref{eq:scenario1} is
 ${\mathbf{w}^{\star}} = \sqrt {\frac{{\delta _{e,1}^2\left( {1 - \frac{1}
{{{2^R}}}} \right)}}
{{\rho \left( \Lambda  \right)}}} {{\mathbf{v}}_{\max }}$,
 where  ${{\mathbf{v}}_{\max }}$ denotes the normalized eigenvector of $\Lambda $
 associated with ${\rho \left( \Lambda  \right)}$.}
   \begin{IEEEproof}
We first show that if $\rho \left( {\Lambda} \right) > 0$ holds true, then problem \eqref{eq:scenario1}  is feasible.
Let ${\mathbf{w}} = l{{\mathbf{v}}_{\max }}$, and we have
\begin{align}
  &{{\mathbf{w}}^H}\left( {{{\mathbf{G}}_1}\ln {p_{k,out}} + \frac{{\delta _{e,1}^2}}
{{\delta _b^2{2^R}}}{\mathbf{h}}{{\mathbf{h}}^H}} \right){\mathbf{w}}
   = {l^2}{\mathbf{v}}_{\max }^H\Lambda {{\mathbf{v}}_{\max }}
   = {l^2}\rho \left( \Lambda  \right).
\end{align}
Since $\rho \left( \Lambda  \right)>0$,  obviously the constraint \eqref{eq:15b} will be satisfied  by increasing $l$. Hence  we can obtain a feasible solution ${\mathbf{w}}$.

Next, we show the reverse direction of the claim, that if problem \eqref{eq:scenario1} is feasible, then $\rho \left( {\Lambda} \right) > 0$ is true.
 If  $\rho \left( \Lambda  \right) \leq 0$,  we have
${{\mathbf{G}}_1}{\text{ln}}{p_{k,out}} + \frac{{\delta _{e,1}^2}}
{{\delta _b^2{2^R}}}{\mathbf{h}}{{\mathbf{h}}^H} \preceq {\mathbf{0}}$.
  Then the left hand side of constraint \eqref{eq:15b} is
\begin{align}
{{\mathbf{w}}^H}\left( {{{\mathbf{G}}_1}{\text{ln}}{p_{k,out}} + \frac{{\delta _{e,1}^2}}
{{\delta _b^2{2^R}}}{\mathbf{h}}{{\mathbf{h}}^H}} \right){\mathbf{w}} \leq  0,~\forall~ {\mathbf{w}}.\label{eq:20}
\end{align}
For $R > 0$, the right hand side of constraint \eqref{eq:15b} is
 \begin{align}
 \delta _{e,1}^2\left( {1 - \frac{1}
{{{2^R}}}} \right) > 0.\label{eq:21}
\end{align}
Hence the constraint \eqref{eq:15b}  cannot hold, which is a contradiction.
Therefore, when $K=1$, the problem \eqref{eq:scenario1} under the  scenario  1 is feasible if and only
 if $\rho \left( \Lambda  \right)>0$.

Finally, we show that the optimal solution can be expressed as ${\mathbf{w}} = \sqrt {\frac{{\delta _{e,1}^2\left( {1 - \frac{1}
{{{2^R}}}} \right)}}
{{\rho \left( \Lambda  \right)}}} {{\mathbf{v}}_{\max }}$.
Consider the following inequality: $  {{\mathbf{w}}^H}\Lambda {\mathbf{w}}
   \leq \rho \left( \Lambda  \right){\left\| {\mathbf{w}} \right\|^2}$,
where the equality is achieved when ${\mathbf{w}}$
 is an eigenvector of
$\Lambda $
 corresponding to the maximum eigenvalue $\rho \left( \Lambda  \right)$.
On the other hand, if constraint \eqref{eq:15b} is satisfied, we must have
$\rho \left( \Lambda  \right){\left\| {\mathbf{w}} \right\|^2} \geq  \delta _{e,1}^2\left( {1 - \frac{1}
{{{2^R}}}} \right)$.
Therefore, the minimum value of the objective function is ${\left\| {\mathbf{w}} \right\|^2} = \frac{{\delta _{e,1}^2\left( {1 - \frac{1}
{{{2^R}}}} \right)}}
{{\rho \left( \Lambda  \right)}}$ and the optimal solution is ${\mathbf{w}}^{\star} = \sqrt {\frac{{\delta _{e,1}^2\left( {1 - \frac{1}
{{{2^R}}}} \right)}}
{{\rho \left( \Lambda  \right)}}} {{\mathbf{v}}_{\max }}$.

\end{IEEEproof}

\textbf{ Proposition 2:}
{\it For scenario 1 with multiple eavesdroppers (i.e., when $K>1$), problem \eqref{eq:scenario1} is feasible if the following holds true
\begin{align}&\frac{{\delta _{e,k}^2}}
{{\delta _b^2{2^R}}}{\left\| {\mathbf{h}} \right\|^4} \geq   - \rho \left( {{{\mathbf{G}}_k}} \right){\left\| {\mathbf{h}} \right\|^2}{\text{ln}}{p_{k,out}} - \delta _{e,k}^2\left( {1 - \frac{1}
{{{2^R}}}} \right), \forall k \in \mathcal{K}.\label{eq:24}
\end{align}
Moreover, if problem \eqref{eq:scenario1} is feasible, then we must have
\begin{align}\rho \left( {{{\mathbf{G}}_k}\ln {p_{k,out}} + \frac{{\delta _{e,k}^2}}
{{\delta _b^2{2^R}}}{\mathbf{h}}{{\mathbf{h}}^H}} \right) > 0, \forall k \in \mathcal{K}.\label{eq:25}\end{align}}
\begin{proof}
Please see Appendix A for proof.
\end{proof}

For $K=1$, we have shown in Proposition 1 that problem \eqref{eq:scenario1} admits closed-form solution.
However, for $K>1$, such closed-form solution is not likely to exist, because general nonconvex QCQP problems are NP-hard \cite{Sidiropoulos}. Fortunately, due to some special structures of problem \eqref{eq:scenario1}, its global optimal solution can still be obtained in polynomial time. In the following, we use the SDR approach for such purpose.

To this end, we first rewrite the problem \eqref{eq:scenario1} equivalently as
\begin{subequations}\label{eq:scenario1_SDP}
\begin{align}
  &\mathop {\min }\limits_{\mathbf{W}} {\text{Tr}}\left( {\mathbf{W}} \right) \label{eq:26a} \\
  {\text{s.t}}&{\text{.Tr}}\left( {\left( {{{\mathbf{G}}_{k}}\ln {p_{k,out}} + \frac{{\delta _{e,k}^2}}
{{\delta _b^2{2^R}}}{\mathbf{h}}{{\mathbf{h}}^H}} \right){\mathbf{W}}} \right) \geq \delta _{e,k}^2\left( {1 - \frac{1}
{{{2^R}}}} \right), \forall k \in \mathcal{K},  \label{eq:26b} \\
 & {\mathbf{W}} \succeq {\mathbf{0}},~{\text{rank}}\left( {\mathbf{W}} \right) = 1.  \label{eq:26c}
\end{align}
\end{subequations}
Dropping the rank constraint, we obtain the following relaxed convex program
\begin{subequations}\label{eq:scenario1_SDR}
\begin{align}
  &\mathop {\min }\limits_{\mathbf{W}} {\text{Tr}}\left( {\mathbf{W}} \right) \label{eq:28a} \\
  {\text{s.t}}&{\text{.Tr}}\left( {\left( {{{\mathbf{G}}_{k}}\ln {p_{k,out}} + \frac{{\delta _{e,k}^2}}
{{\delta _b^2{2^R}}}{\mathbf{h}}{{\mathbf{h}}^H}} \right){\mathbf{W}}} \right) \geq \delta _{e,k}^2\left( {1 - \frac{1}
{{{2^R}}}} \right), \forall k \in \mathcal{K}, \label{eq:28b} \\
 & {\mathbf{W}} \succeq {\mathbf{0}}\label{eq:28c},
\end{align}
\end{subequations}
whose optimal solution can be efficiently obtained by interior-point algorithms \cite{Sturm,Grant}.
Generally speaking, there is a positive gap between the optimal objective value of the original problem and its rank-relaxed counterpart, as there is no guarantee that the solution for the relaxed problem is of rank one. However, below we show that in our case, the solution of \eqref{eq:scenario1_SDR} is indeed of rank one. That is, there is no loss of optimality in performing the relaxation.

\textbf{Theorem 1:}  {\it Suppose $R>0$, and that problem \eqref{eq:scenario1_SDR} is feasible. Then the
optimal solution of the problem \eqref{eq:scenario1_SDR} must be of rank one.}
\begin{proof}
Please see Appendix B for proof.
\end{proof}

\subsection{Scenario 2: Perfect LCSI  and imperfect ECSI}

In this subsection, we solve the power minimization problem \eqref{eq:power_min} under the assumption of perfect LCSI and imperfect  ECSI. The main approach we will employ is a relaxation-restriction procedure\footnote{\footnotesize {Similar relaxation-restriction procedure was also used in \cite{Wang_b}, but for the purpose of handling the outage constrained MISO downlink beamformer design problem. In contrast, in our work we apply the procedure to solve the outage  constrained  secure transmission  problem with different  the Bernstein-type Inequality to handle the chance constraints. Furthermore, we propose the Projection Approximation  Procedure to tackle  non rank-one solution case.  }}: we first perform an SDR to lift the problem into a high dimension (the relaxation step), and then conservatively transform the resulting chance constraint into a deterministic form (the restriction step).

\subsubsection{Semidefinite Relaxation}
 We first reformulate the chance constraint in \eqref{eq:11b}. Specifically, the inequality
\begin{align}{\log _2}\left( {1 + \frac{{{{\mathbf{w}}^H}{\mathbf{h}}{{\mathbf{h}}^H}{\mathbf{w}}}}
{{\delta _b^2}}} \right) - {\log _2}\left( {1 + \frac{{{{\mathbf{w}}^H}{{\mathbf{g}}_k}{\mathbf{g}}_k^H{\mathbf{w}}}}
{{\delta _{e,k}^2}}} \right) \geq  R\label{eq:29}
\end{align}
 can be rewritten as
\begin{align}{2^{ - R}}\delta _{e,k}^2\left( {\delta _b^2 + {{\mathbf{w}}^H}{\mathbf{h}}{{\mathbf{h}}^H}{\mathbf{w}}} \right) \geq  \delta _b^2\left( {\delta _{e,k}^2 + {{\mathbf{w}}^H}{{\mathbf{g}}_k}{\mathbf{g}}_k^H{\mathbf{w}}} \right),\label{eq:30}
\end{align}
which can be further rewritten as
\begin{align}{2^{ - R}}\delta _{e,k}^2\left( {\delta _b^2 + {{\mathbf{h}}^H}{\mathbf{w}}{{\mathbf{w}}^H}{\mathbf{h}}} \right) \geq  \delta _b^2\left( {\delta _{e,k}^2 + {\mathbf{g}}_k^H{\mathbf{w}}{{\mathbf{w}}^H}{{\mathbf{g}}_k}} \right)\label{eq:31}.
\end{align}
Define $\bW\triangleq\bw\bw^H$, and plug in  the definition of imperfect ECSI \eqref{eq:7} in \eqref{eq:31}, we obtain
\begin{align}
  &\delta _{e,k}^2 + \Delta {\mathbf{g}}_k^H{\mathbf{W}}\Delta {{\mathbf{g}}_k} + 2\operatorname{Re} \left( {\Delta {\mathbf{g}}_k^H{\mathbf{W}}{{\widehat{\mathbf{g}}}_k}} \right) + \widehat{\mathbf{g}}_k^H{\mathbf{W}}{\widehat{\mathbf{g}}_k} \leq  \frac{{{2^{ - R}}\delta _{e,k}^2}}
{{\delta _b^2}}\left( {\delta _b^2 + {{\mathbf{h}}^H}{\mathbf{Wh}}} \right). \label{eq:32}
\end{align}

It follows that problem \eqref{eq:power_min}  can be  equivalently  reformulated as
  \begin{subequations}\label{eq:scenario2}
\begin{align}
  &\mathop {\min }\limits_{\mathbf{W}} {\text{Tr}}\left( {\mathbf{W}} \right) \label{eq:33a} \\
  {\text{s.t.}}&\Pr \Bigg{\{} {\Delta {\mathbf{g}}_k^H{\mathbf{W}}\Delta {{\mathbf{g}}_k} + 2\operatorname{Re} \left( {\Delta {\mathbf{g}}_k^H{\mathbf{W}}{{\widehat{\mathbf{g}}}_k}} \right) + \widehat{\mathbf{g}}_k^H{\mathbf{W}}{{\widehat{\mathbf{g}}}_k}}{ - \frac{{{2^{ - R}}\delta _{e,k}^2}}
{{\delta _b^2}}\left( {\delta _b^2 + {{\mathbf{h}}^H}{\mathbf{Wh}}} \right) + \delta _{e,k}^2 \geq 0} \Bigg{\}} \nonumber \\
  &\leq {p_{k,out}},  ~\forall k \in \mathcal{K},  \label{eq:33b}\\
  &{\mathbf{W}} \succeq {\mathbf{0}},{\text{rank}}\left( {\mathbf{W}} \right) = 1. \label{eq:33c}
\end{align}
\end{subequations}

Using the SDR approach, we relax problem \eqref{eq:scenario2} by again dropping the rank constraint ${\text{rank}}\left( {\mathbf{W}} \right) = 1$. The rank relaxed problem becomes
\begin{align}
  &\mathop {{\text{min}}}\limits_{{\mathbf{W}} } ~{\text{Tr}}\left( {\mathbf{W}} \right) \label{eq:34} \\
  \text{s.t.}&~\eqref{eq:33b},~{\mathbf{W}} \succeq \mathbf{0} \nonumber.
\end{align}

Observe that the constraint \eqref{eq:33b} is still a difficult chance
constraint. In the following, we transform such chance constraint into a
deterministic form by utilizing the Bernstein-type inequality I \cite{Bechar,Ma2}.

\subsubsection{Conservative Transformation}  Let us rewrite the
CSI error  as  $\Delta {{\mathbf{g}}_k} = {\mathbf{E}}_{e,k}^{1/2}{{\mathbf{x}}_{e,k}}$
 where  ${{\mathbf{x}}_{e,k}} \sim \mathcal{C}\mathcal{N}\left( {{\mathbf{0}},{\mathbf{I}}} \right)$.
   Then, the chance constraint
 \eqref{eq:33b}  can be represented  as follows
  \begin{align}
 &\Pr \left\{ {{\bf{x}}_{e,k}^H{{\bf{A}}_{e,k}}{{\bf{x}}_{e,k}} + 2{\mathop{\rm Re}\nolimits} \left\{ {{\bf{x}}_{e,k}^H{{\bf{a}}_{e,k}}} \right\}  \ge {c_{e,k}}} \right\} \le {{p_{k,out}}},~\forall k \in \mathcal{K}, \label{eq:35}
\end{align}
where we have defined $
{{\mathbf{A}}_{e,k}}  \triangleq {\mathbf{E}}_{e,k}^{1/2}{\mathbf{WE}}_{e,k}^{1/2}$,
${{\mathbf{a}}_{e,k}}  \triangleq {\mathbf{E}}_{e,k}^{1/2}{\mathbf{W}}{\widehat{\mathbf{g}}_k}$,
and ${c_{e,k}} \triangleq \frac{{{2^{ - R}}\delta _{e,k}^2}}
{{\delta _b^2}}\left( {\delta _b^2 + {{\mathbf{h}}^H}{\mathbf{Wh}}} \right) - \widehat{\mathbf{g}}_k^H{\mathbf{W}}{\widehat{\mathbf{g}}_k} - \delta _{e,k}^2$.

The Bernstein-type inequality I, stated below, is used to bound the tail probability of quadratic forms
of Gaussian variables involving matrices.

\textbf{Lemma 1 (The Bernstein-type Inequality I)} \cite{Bechar,Ma2} {\it Let ${\mathbf{G}}
 = {{\mathbf{x}}^H}{\mathbf{Ax}} + 2\operatorname{Re} \left\{ {{{\mathbf{x}}^H}{\mathbf{a}}} \right\}$,
 where ${\mathbf{A}} \in {\mathbb{C}^{N \times N}}$ is a complex hermitian
 matrix, $\mathbf{{a}} \in {\mathbb{C}^{{N}}}$, and ${\mathbf{{x}}} \sim \mathcal{CN}\left( {\mathbf{0},\mathbf{I}}
 \right)$. Then for any $\sigma  \geq 0$, we have
 \begin{align}
  &\Pr \left\{ {\mathbf{{G}} \geq {\text{Tr}}\left( \mathbf{{A}} \right) + \sqrt {2\sigma } \sqrt {{{\left\| {{\text{vec}}\left( \mathbf{{A}} \right)} \right\|}^2} + 2{{\left\| \mathbf{{a}} \right\|}^2}}  + \sigma {s^ + }\left( \mathbf{{A}} \right)} \right\}\leq \exp ( - \sigma ), \label{eq:36}
  \end{align}
where   ${s^ + }\left( {\mathbf{{A}}} \right) = \max \left\{ {{\lambda
_{\max }}\left( {\mathbf{{A}}} \right),0} \right\}$
 with
${{\lambda _{\max }}\left( {{\mathbf{{A}}}} \right)}$ denotes the
maximum eigenvalue of matrix ${{\mathbf{{A}}}}$.}

With the Bernstein-type inequality I, the chance constraint \eqref{eq:35}  can be conservatively transformed into the following deterministic form:
 \begin{align}
  &{\text{Tr}}\left( {{{\mathbf{A}}_{e,k}}} \right) + \sqrt {2{\sigma _{e,k}}} \sqrt {{{\left\| {{\text{vec}}\left( {{{\mathbf{A}}_{e,k}}} \right)} \right\|}^2} + 2{{\left\| {{{\mathbf{a}}_{e,k}}} \right\|}^2}}+ {\sigma _{e,k}}{s^ + }\left( {{{\mathbf{A}}_{e,k}}} \right) - {c_{e,k}} \leq 0,~\forall k \in \mathcal{K}, \label{eq:37}
\end{align}
where  ${\sigma _{e,k}} =  - \ln \left( { {p_{k,out}}} \right)$. That is, if \eqref{eq:37} is true, then the chance constraint \eqref{eq:35} must hold true.
Consequently, the relaxed problem \eqref{eq:34} is now
conservatively reformulated as
\begin{align}
 &\mathop {{\text{min}}}\limits_{{\mathbf{W}}  } ~{\text{Tr}}\left( {\mathbf{W}} \right)   \label{eq:39} \\
  {\text{s}}.{\text{t}}.\quad &\eqref{eq:37},~{{\mathbf{{W}}}} \succeq \mathbf{0} \nonumber.
  \end{align}

It is easy to see that the above problem is equivalent to the following problem
 \begin{align}
  &\mathop {{\text{min}}}\limits_{{\mathbf{W}}  } ~{\text{Tr}}\left( {\mathbf{W}} \right)   \label{eq:scenario2_SDP} \\
  {\text{s}}.{\text{t}}.\quad
  &  {\text{Tr}}\left( {{{\mathbf{A}}_{e,k}}} \right) + \sqrt {2{\sigma _{e,k}}} {\mu _{e,k}} + {\sigma _{e,k}}{v_{e,k}} - {c_{e,k}} \leq  0, \ \forall k \in \mathcal{K},  \hfill \nonumber\\
  &\quad \left\| {\begin{array}{*{20}{c}}
   {{\text{vec}}\left( {{{\mathbf{A}}_{e,k}}} \right)}  \\
   {\sqrt 2 {{\mathbf{a}}_{e,k}}}  \\
 \end{array} } \right\| \leq  {\mu _{e,k}}, \ \forall k \in \mathcal{K},  \hfill \nonumber\\
  &\quad {v_{e,k}}{\mathbf{I}} - {{\mathbf{A}}_{e,k}} \succeq {\mathbf{0}}, \quad {v_{e,k}} \geq  0 \forall k \in \mathcal{K},  \hfill \nonumber\\
  &\quad \bW\succeq 0, \nonumber
\end{align}
where ${\mu_{e,k}}$ and ${v_{e,k}}$, $\forall k \in \mathcal{K}$,
are slack variables. This problem has a linear objective, and it includes $K$ linear constraints, $K$ second order cone constraints and  $K+1$  convex PSD constraints. Therefore it is a convex problem and can be solved
by using off-the-shelf convex optimization solvers, such as CVX \cite{Grant}. However, due to the rank relaxation, there is no guarantee that the resulting optimal solution ${{\mathbf{W}}^{{\text{opt}}}}$ is feasible for the original problem \eqref{eq:scenario2}. To obtain a feasible rank-one solution $\bw^*$, we propose a simple Projection Approximation  Procedure, which is summarized in Algorithm 1. Surprisingly, this simple scheme is guaranteed to find a rank-1 solution which has performance no worse than $\mathbf{W}^{\rm opt}$.

\textbf{ Proposition 3:}
{\it
Let $\bW^{\rm opt}$ denote the optimal solution of problem \eqref{eq:scenario2_SDP}.  If $\rank({\mathbf{W}}^{{\text{opt}}})>1$, then the  Projection Approximation  Procedure can provide a rank-one solution $\widehat{\mathbf{W}}$ with the following performance guarantee: ${\text{Tr}}\left( {\widehat{\mathbf{W}}} \right) \leq  {\text{Tr}}\left( {{{\mathbf{W}}^{opt}}} \right)$.
}
\begin{proof}
Please see Appendix C for proof.
\end{proof}

\begin{figure}[htbp]
\centering
\includegraphics[width=6cm]{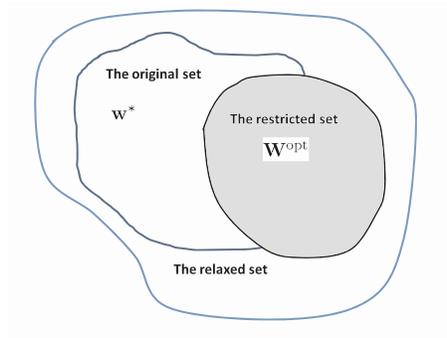}
\caption{Graphical representation of the relaxation restriction procedure.}
\label{fig_rr}
\end{figure}

\textbf{Remark 2} (The relaxation-restriction procedure): The feasible region for the original probability constrained problem \eqref{eq:scenario2} is nonconvex. The relaxation step expands the feasible region to a larger, albeit still nonconvex set. By using the Bernstein-type inequality I, the latter set shrinks to a convex set (the shaded region in Fig. \ref{fig_rr}), defined by Eq. \eqref{eq:37}, thus the restricted problem becomes a convex one \eqref{eq:scenario2_SDP}. Proposition 3 states that, remarkably, the Projection Approximation Procedure is able to find a feasible solution $\bw^*$ in the {\it original} feasible region that is at least as good as the solution $\bW^{\rm opt}$ of the restricted convex problem \eqref{eq:scenario2_SDP}.

\begin{table}[t]
 \begin{tabular}{l}
\hline Algorithm 1: Projection Approximation  Procedure
\\
1.Let ${\mathbf{P}}$ denote the project matrix of vector ${{\mathbf{W}}^{{1 \mathord{\left/
 {\vphantom {1 2}} \right.
 \kern-\nulldelimiterspace} 2}}}{\mathbf{h}}$, where
  $\begin{array}{*{20}{c}}
   {{\mathbf{P}} = \frac{{{{\mathbf{W}}^{{1 \mathord{\left/
 {\vphantom {1 2}} \right.
 \kern-\nulldelimiterspace} 2}}}{\mathbf{h}}{{\left( {{{\mathbf{W}}^{{1 \mathord{\left/
 {\vphantom {1 2}} \right.
 \kern-\nulldelimiterspace} 2}}}{\mathbf{h}}} \right)}^H}}}
{{{{\left\| {{\mathbf{h}}{{\mathbf{W}}^{{1 \mathord{\left/
 {\vphantom {1 2}} \right.
 \kern-\nulldelimiterspace} 2}}}} \right\|}^2}}}} \hfill  \\
 \end{array}$;\\
 2. We construct a new rank one   solution
 $ \widehat{\mathbf{W}} = {{\mathbf{W}}^{{1 \mathord{\left/
 {\vphantom {1 2}} \right.
 \kern-\nulldelimiterspace} 2}}}{\mathbf{P}}{{\mathbf{W}}^{{1 \mathord{\left/
 {\vphantom {1 2}} \right.
 \kern-\nulldelimiterspace} 2}}}$;\\
 3. By SVD method, we can obtain ${\mathbf{w}}^*$ from $\widehat{\mathbf{W}}$.\\
  \hline
\end{tabular}
\label{alg:1}
\end{table}



\subsection{Scenario 3: Imperfect LCSI  and imperfect ECSI}

In this subsection, we discuss problem \eqref{eq:power_min} when the knowledge of both LCSI and ECSI are imperfect. 
Note that in this scenario, {\it multiple types} of independent CSI errors are included in each chance constraint. The resulting problem is different, and arguably more difficult, compared with the problem considered in the previous scenario, where each constraint involves only {\it a single type} of CSI error. Our main approach is again the relaxation-restriction procedure used in the previous subsection. However, in the restriction step a different form of Bernstein-type inequality needs to be used.

 \subsubsection{Semidefinite Relaxation}

 Using the imperfect CSI model \eqref{eq:8},  problem \eqref{eq:power_min}  can be  equivalently  reformulated as:
 \begin{subequations}\label{eq:scenario3_SDP}
\begin{align}
  &\mathop {{\text{min}}}\limits_{\mathbf{W}} {\text{Tr}}\left( {\mathbf{W}} \right) \label{eq:40a}\\
  {\text{s}}{\text{.t}}{\text{.}}&\Pr \left( {\left[ {\Delta {{\mathbf{h}}^H},\Delta {\mathbf{g}}_k^H} \right]{\text{diag}}\left\{ {\frac{1}
{{\delta _n^2}}{\mathbf{W}}, - \frac{{{2^R}}}
{{\delta _{e,k}^2}}{\mathbf{W}}} \right\}{{\left[ {\Delta {{\mathbf{h}}^H},\Delta {\mathbf{g}}_k^H} \right]}^H}} \right. \nonumber \\
   &+ 2{\text{Re}}\left\{ {\left[ {\Delta {{\mathbf{h}}^H},\Delta {\mathbf{g}}_k^H} \right]{\text{diag}}\left\{ {\frac{1}
{{\delta _n^2}}{\mathbf{W}}, - \frac{{{2^R}}}
{{\delta _{e,k}^2}}{\mathbf{W}}} \right\}{{\left[ {{{\widehat{\mathbf{h}}}^H},\widehat{\mathbf{g}}_k^H} \right]}^H}} \right\} \nonumber \\
  &\left. { + \left[ {{{\widehat{\mathbf{h}}}^H},\widehat{\mathbf{g}}_k^H} \right]{\text{diag}}\left\{ {\frac{1}
{{\delta _n^2}}{\mathbf{W}}, - \frac{{{2^R}}}
{{\delta _{e,k}^2}}{\mathbf{W}}} \right\}{{\left[ {{{\widehat{\mathbf{h}}}^H},\widehat{\mathbf{g}}_k^H} \right]}^H} \geq  {2^R} - 1} \right) \geq 1 - {p_{out}},\forall k \in \mathcal{K}, \label{eq:40b} \\
  &{\mathbf{W}} \succeq {\mathbf{0}},{\text{rank}}\left( {\mathbf{W}} \right) = 1, \label{eq:40c}
\end{align}
\end{subequations}
where ${\mathbf{W}} = {\mathbf{w}}{{\mathbf{w}}^H}$.
Again we obtain the following relaxed problem of \eqref{eq:scenario3_SDP}, by dropping the rank constraint:
\begin{align}
  &\mathop {{\text{min}}}\limits_{{\mathbf{W}} } ~{\text{Tr}}\left( {\mathbf{W}} \right) \label{eq:scenario3_SDR} \\
  \text{s.t.}&~\eqref{eq:40b},~{\mathbf{W}} \succeq \mathbf{0} \nonumber.
\end{align}

Next we transform the chance constraint \eqref{eq:40b} into a
deterministic form. To this end, let us rewrite the
CSI error  as $\Delta {\mathbf{h}} = {\mathbf{E}}_b^{1/2}{{\mathbf{x}}_h}$, and $\Delta {{\mathbf{g}}_k} = {\mathbf{E}}_{e,k}^{1/2}{{\mathbf{x}}_{e,k}}$
 where  ${{\mathbf{x}}_{{{\mathbf{h}}}}} \sim \mathcal{C}\mathcal{N}\left( {{\mathbf{0}},{\mathbf{I}}} \right)$ and ${{\mathbf{x}}_{e,k}} \sim \mathcal{C}\mathcal{N}\left( {{\mathbf{0}},{\mathbf{I}}} \right)$. Further define ${\widetilde{\mathbf{x}}_k} \triangleq {\left[ {{\mathbf{x}}_h^H,{\mathbf{x}}_{e,k}^H} \right]^H}$, $\forall k \in \mathcal{K}$.
  Then, the chance constraint \eqref{eq:40b} can be written as
  \begin{align}
 &\Pr \left( {\widetilde{\mathbf{x}}_k^H{{\mathbf{A}}_k}{{\widetilde{\mathbf{x}}}_k} + \widetilde{\mathbf{x}}_k^H{{\mathbf{a}}_k} \leq {c_k}} \right) \leq {p_{{out}}},\forall k \in \mathcal{K}  \label{eq:42}
\end{align}
where
${{\mathbf{A}}_k} \triangleq {\text{diag}}\left\{ {\frac{1}
{{\delta _n^2}}{\mathbf{E}}_s^{1/2}{\mathbf{WE}}_s^{1/2}, - \frac{{{2^{{R}}}}}
{{\delta _{e,k}^2}}{\mathbf{E}}_{e,k}^{1/2}{\mathbf{WE}}_{e,k}^{1/2}} \right\}$,
${{\mathbf{a}}_k} \triangleq {\text{diag}}\left\{ {\frac{1}
{{\delta _n^2}}{\mathbf{E}}_s^{1/2}{\mathbf{W}}, - \frac{{{2^R}}}
{{\delta _{e,k}^2}}{\mathbf{E}}_{e,k}^{1/2}{\mathbf{W}}} \right\}{\left[ {{{\widehat{\mathbf{h}}}^H},\widehat{\mathbf{g}}_k^H} \right]^H}$,
 and
${c_k} \triangleq {2^R} - \left[ {{{\widehat{\mathbf{h}}}^H},\widehat{\mathbf{g}}_k^H} \right]{\text{diag}}\left\{ {\frac{1}
{{\delta _n^2}}{\mathbf{W}}, - \frac{{{2^R}}}
{{\delta _{e,k}^2}}{\mathbf{W}}} \right\}{\left[ {{{\widehat{\mathbf{h}}}^H},\widehat{\mathbf{g}}_k^H} \right]^H} - 1$.

It is worth noting that constraint \eqref{eq:42} takes a different form from \eqref{eq:35}. Thus we will need a different type of Bernstein inequality to transform this constraint.

\textbf{ Lemma $2$ (The Bernstein-type Inequality II)} \cite{Bechar,Wang_a,Wang_b,Ma2} {\it Let ${\mathbf{G}}
 = {{\mathbf{x}}^H}{\mathbf{Ax}} + 2\operatorname{Re} \left\{ {{{\mathbf{x}}^H}{\mathbf{a}}} \right\}$,
 where ${\mathbf{A}} \in {\mathbb{C}^{N \times N}}$ is a complex hermitian
 matrix, $\mathbf{{a}} \in {\mathbb{C}^{{N}}}$, and ${\mathbf{{x}}} \sim \mathcal{CN}\left( {\mathbf{0},\mathbf{I}}
 \right)$. Then for any $\sigma  \geq 0$, we have
 \begin{align}
  &\Pr \left\{ {\mathbf{{G}} \leq {\text{Tr}}\left( \mathbf{{A}} \right) - \sqrt {2\sigma } \sqrt {{{\left\| {{\text{vec}}\left( \mathbf{{A}} \right)} \right\|}^2} + 2{{\left\| \mathbf{{a}} \right\|}^2}}  - \sigma {s^ - }\left( \mathbf{{A}} \right)} \right\}\leq \exp ( - \sigma ),   \label{eq:43}
   \end{align}
where  ${s^ - }\left( {\mathbf{{A}}} \right) = \max \left\{
{{\lambda _{\max }}\left( { - {\mathbf{{A}}}} \right),0} \right\}$
 with
${{\lambda _{\max }}\left( {{\mathbf{{A}}}} \right)}$ denotes the
maximum eigenvalue of matrix ${{\mathbf{{A}}}}$.}

With the Bernstein-type inequality II, the chance
constraint \eqref{eq:42} can be conservatively transformed into the following  deterministic form:
 \begin{align}
 &  {\text{Tr}}\left( {{{\mathbf{A}}_k}} \right) - \sqrt {2{\sigma_k}} \sqrt {{{\left\| {{\text{vec}}\left( {{{\mathbf{A}}_k}} \right)} \right\|}^2} + 2{{\left\| {{{\mathbf{a}}_k}} \right\|}^2}}   - {\sigma_k  }{s^ - }\left( {{{\mathbf{A}}_k}} \right) - {c_k} \geq  0, \forall k \in \mathcal{K},\label{eq:44}
 \end{align}
where ${\sigma_k } =  - \ln \left( {{p_{k,out}}} \right)$.

 That is, if \eqref{eq:44} is true, then the chance constraint \eqref{eq:42} must hold true.
Consequently, the relaxed problem \eqref{eq:scenario3_SDR} is now
conservatively reformulated as
\begin{align}
 &\mathop {{\text{min}}}\limits_{{\mathbf{W}}  } ~{\text{Tr}}\left( {\mathbf{W}} \right)   \label{eq:45} \\
  {\text{s}}.{\text{t}}.\quad &\eqref{eq:44},~{{\mathbf{{W}}}} \succeq \mathbf{0} \nonumber.
  \end{align}
which is equivalent to
 \begin{align}
  &\mathop {{\text{min}}}\limits_{\mathbf{W}} {\text{Tr}}\left( {\mathbf{W}} \right) \label{eq:scenario3_final} \\
  {\text{s}}.{\text{t}}.\quad & {\text{Tr}}\left( {{{\mathbf{A}}_k}} \right) - \sqrt {2{\sigma _k}} {\mu _k} - {\sigma _k}{v_k} - {c_k} \geq 0,\forall k \in \mathcal{K}, \nonumber \\
  &\left\| {\begin{array}{*{20}{c}}
   {{\text{vec}}\left( {{{\mathbf{A}}_k}} \right)}  \\
   {\sqrt 2 {{\mathbf{a}}_k}}  \\
 \end{array} } \right\| \leq  {\mu _k},~\forall k \in \mathcal{K}, \nonumber \\
  &{v_k}{\mathbf{I}} + {{\mathbf{A}}_k} \succeq {\mathbf{0}},~{v_k} \geq  0~\forall k \in \mathcal{K}, \nonumber \\
    &{{\mathbf{{W}}}} \succeq \mathbf{0} \nonumber.
  \end{align}
  Note that the constraints of problem \eqref{eq:scenario3_final}  includes  linear constraints,   second order cone
 constraints and convex PSD constraints. Thus, problem \eqref{eq:scenario3_final}  is  convex  and can be efficiently solved by existing convex optimization solvers. However,
the resulting optimal solution ${{\mathbf{W}}^{{\text{opt}}}}$ may not
be of rank-one. If this happens, then the well-known Gaussian Randomization Procedure \cite{Luo} can be applied to obtain a feasible solution to problem \eqref{eq:scenario3_final}.

\begin{figure}[htbp]
\centering
\includegraphics[width=8cm]{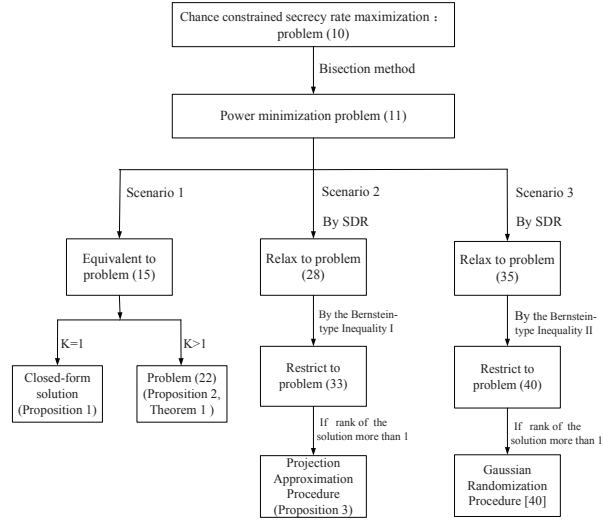}
\caption{The schematic diagram.}
\label{fig:algorithmic}
\end{figure}

In Fig. \ref{fig:algorithmic}, we briefly summarize the algorithmic steps for solving the outage probability constrained secrecy rate maximization problem \eqref{eq:chance_constrained} for all three scenarios considered in this work.

\section{Simulation Results and Discussions}

To illustrate the performance of the schemes proposed in Section III, we present detailed numerical results for all three scenarios. The results to be presented in this section are based on
the following simulation settings (unless otherwise specified): the number of transmit antennas at Alice  is
 $N_t = 6$ and
the noise variance at all receive nodes are the same, i.e., ${\delta _b}={\delta _k}=1$, $\forall k \in \mathcal{K} $.
  The outage probabilities are ${p_{k,out}} = {\overline p _{out}}$, $\forall k \in \mathcal{K} $
 and
${\overline p _{out}} = 0.05$. The average transmit power is $20$dB.

\subsection{Simulation Results for Scenario 1}

 In Scenario 1, Alice knows the full LCSI of ${\mathbf{{h}}}$, but only the statistical ECSI ${{\mathbf{g}}_k}$, $\forall k \in \mathcal{K}$.  In our experiments,
 all channels are in Rayleigh flat fading, i.e., ${\mathbf{{h}}} \sim \mathcal{CN}\left(
{\mathbf{0},\mathbf{I}} \right)$, and the channels of the  Eves to the Alice are different ${{\mathbf{g}}_k} \sim \mathcal{C}\mathcal{N}\left( {{\mathbf{0}},{\varepsilon _e} \times {{\overline {\mathbf{G}} }_k}} \right)$, $\forall k \in \mathcal{K}$.
where
${\overline {\mathbf{G}} _1} = {\mathbf{I}}$, ${\overline {\mathbf{G}} _2} = {\text{diag}}\left\{ {2,1,1,1,1} \right\}$, and ${\overline {\mathbf{G}} _3} = {\text{diag}}\left\{ {1,1,1,1,0.5} \right\}$, and $\varepsilon _e>0$ denotes the value of the ECSI errors variance.

\begin{figure}[h]
    \begin{minipage}[b]{0.45\textwidth}
      \centering
      \includegraphics[height=5cm,width=7cm]{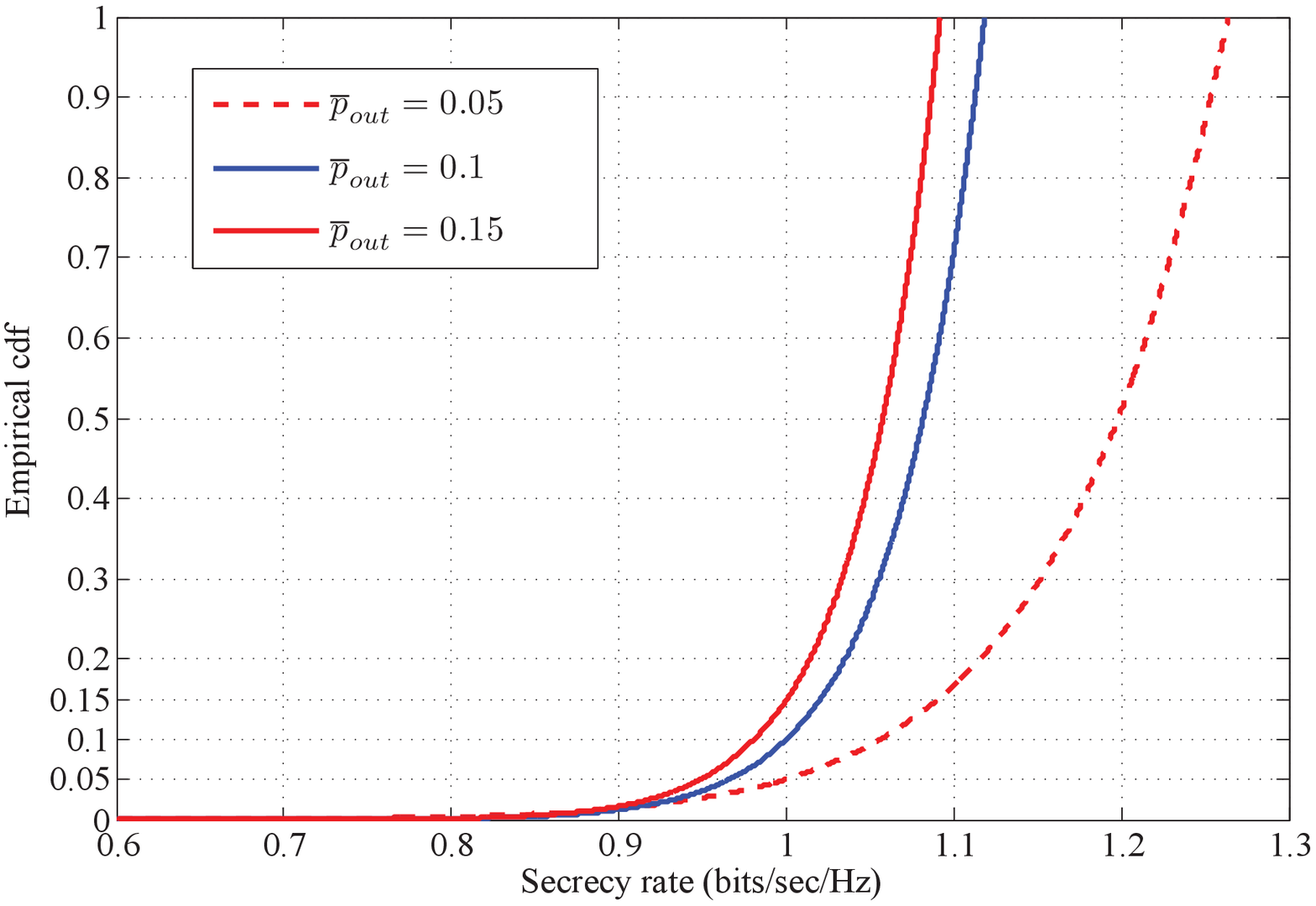}
      \vskip-0.2cm\centering {\footnotesize (a)}
    \end{minipage}\hfill
    \begin{minipage}[b]{0.45\textwidth}
      \centering
      \includegraphics[height=5cm,width=7cm]{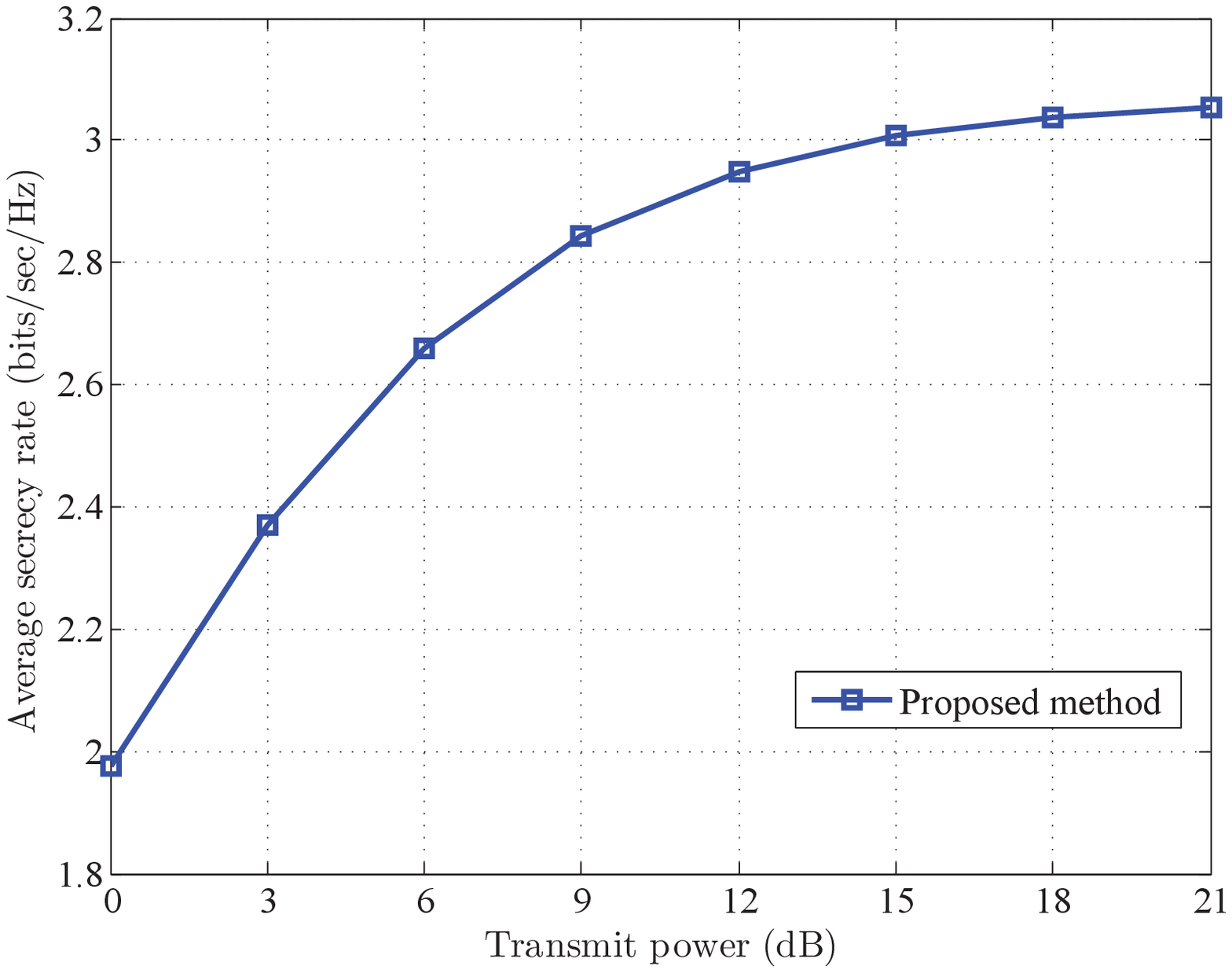}
      \vskip-0.2cm\centering {\footnotesize (b)}
    \end{minipage}\hfill
    \begin{minipage}[b]{0.45\textwidth}
      \centering
      \includegraphics[height=5cm,width=7cm]{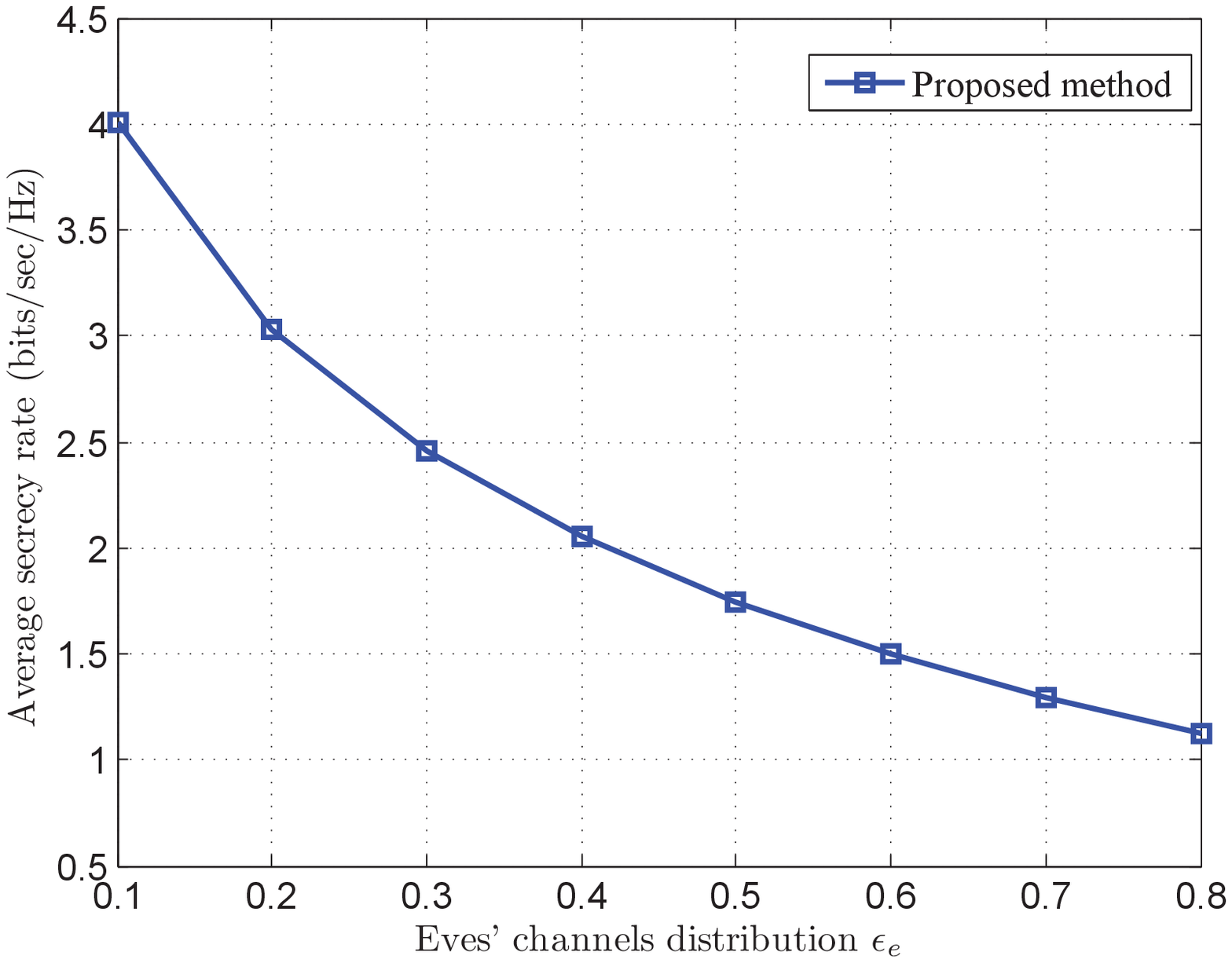}
      \vskip-0.2cm\centering {\footnotesize (c)}
    \end{minipage}
 \caption{(a)~The empirical CDF of  secrecy rate with $R=1$ (bits/sec/Hz) and ${\varepsilon _e} = 0.2$; (b)~Average secrecy rate  versus transmit power for ${\varepsilon _e} = 0.2$; (c)~Average secrecy rate  versus Eves' channels distribution $\varepsilon_e$. }
  \label{fig: Scenario_1} 
\end{figure}

In our first experiment, we demonstrate the robustness of the proposed design. Fig. \ref{fig: Scenario_1} (a) plots the empirical cumulative distribution function (CDF) of the secrecy rates achieved by solving problem \eqref{eq:scenario1}. Each curve in the figure represents the empirical CDF of the secrecy rates obtained from $10000$ random channel realizations. We set the target secrecy rate as $R=1$ (bits/sec/Hz),  the ECSI variance as
${\varepsilon _e} = 0.2$, and used different outage probabilities ${\overline p _{out}} = [0.05,\ 0.1,\ 0.15]$.
 From the figure, we observe that for all three cases simulated, the secrecy rates generated by the proposed method  satisfy the required outage probabilities constraint. Fig. \ref{fig: Scenario_1} (b)  depicts the  achieved secrecy rate  versus  the total transmit power, for the case where the ECSI variance is given by ${\varepsilon _e} = 0.2$ (each point on the figure is the averaged rate over 1000 random channel realizations).  As is shown in this figure, the  secrecy rate improves with  increased transmit power ${\overline P _R}$, but the rate of such improvement decreases. This is because in the high transmit power region, the secrecy rate is limited by what can be achieved by of Eves' channels. On the other hand, Fig. \ref{fig: Scenario_1} (c) shows that the average secrecy rate decreases when the ECSI variance becomes larger.

%

%

\subsection{Simulation Results for Scenario 2}

  Assume that
 all channels are in Rayleigh flat fading, i.e., ${{\mathbf{{h}}}} \sim \mathcal{CN}\left(
{\mathbf{0},\mathbf{I}} \right)$,   and
${\widehat{\mathbf{g}}_k} \sim \mathcal{C}\mathcal{N}\left( {{\mathbf{0}},{\mathbf{I}}} \right)$, $\forall k \in \mathcal{K} $. The variance  of  ECSI  error is ${{\mathbf{E}}_{e,k}} = {\varepsilon _e} \times {\mathbf{I}}$, $\forall k \in \mathcal{K} $,  where the parameter  $\varepsilon _e  \geq  0$ represents the ECSI error variance.
The performance of the proposed design is compared against the worst case design method \cite{Li3}. For a fair comparison, we apply the evaluation methods \cite{Li4,Chung,Pascual,Zheng}, to obtain the upper bound of the CSI error covariance for the worst case design method [19,Proposition 4]. We also present the non-robust method [19, Problem 16] for comparison.



\begin{figure}[h]
    \begin{minipage}[b]{0.45\textwidth}
      \centering
      \includegraphics[height=5cm,width=7cm]{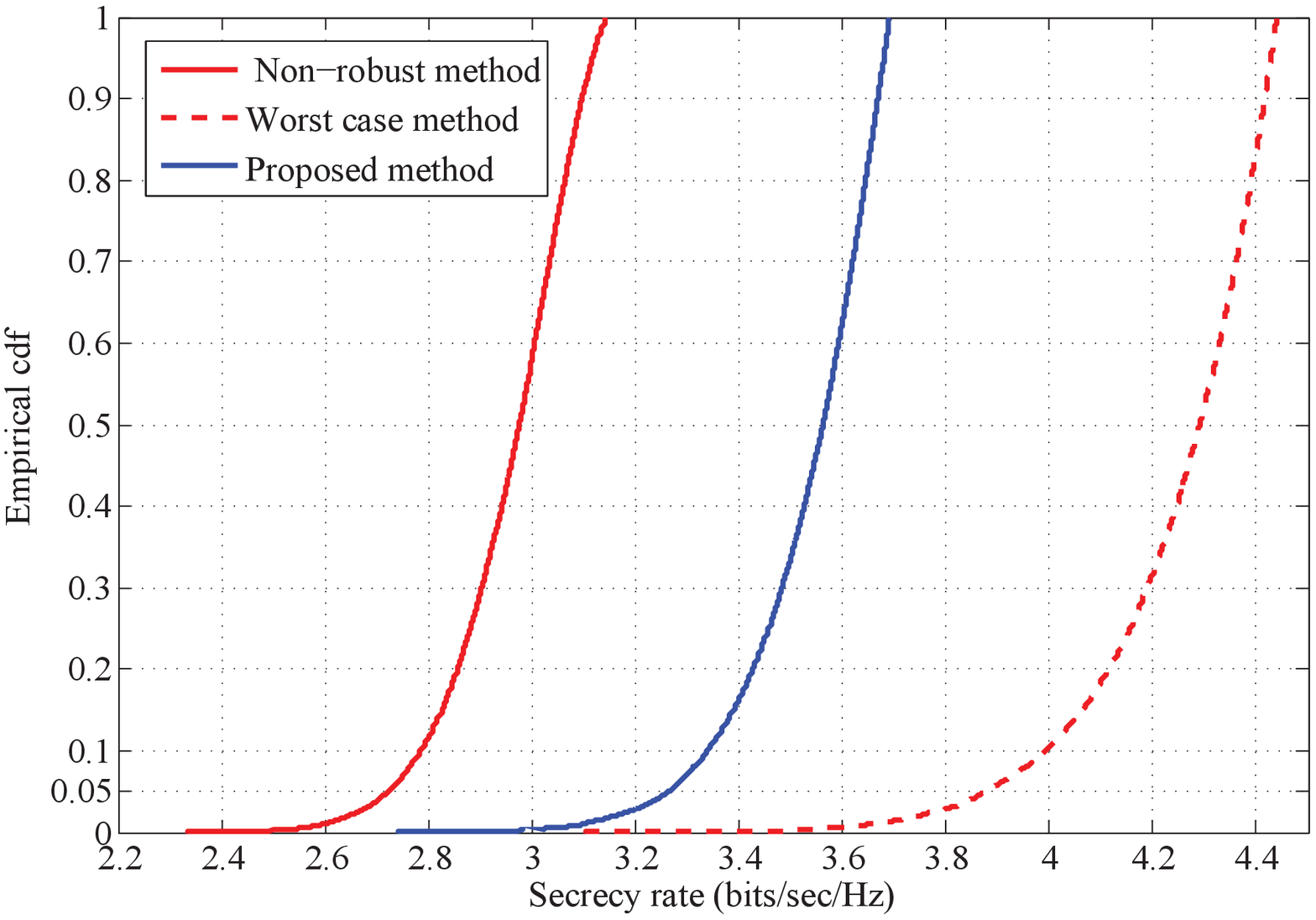}
      \vskip-0.2cm\centering {\footnotesize (a)}
    \end{minipage}\hfill
    \begin{minipage}[b]{0.45\textwidth}
      \centering
      \includegraphics[height=5cm,width=7cm]{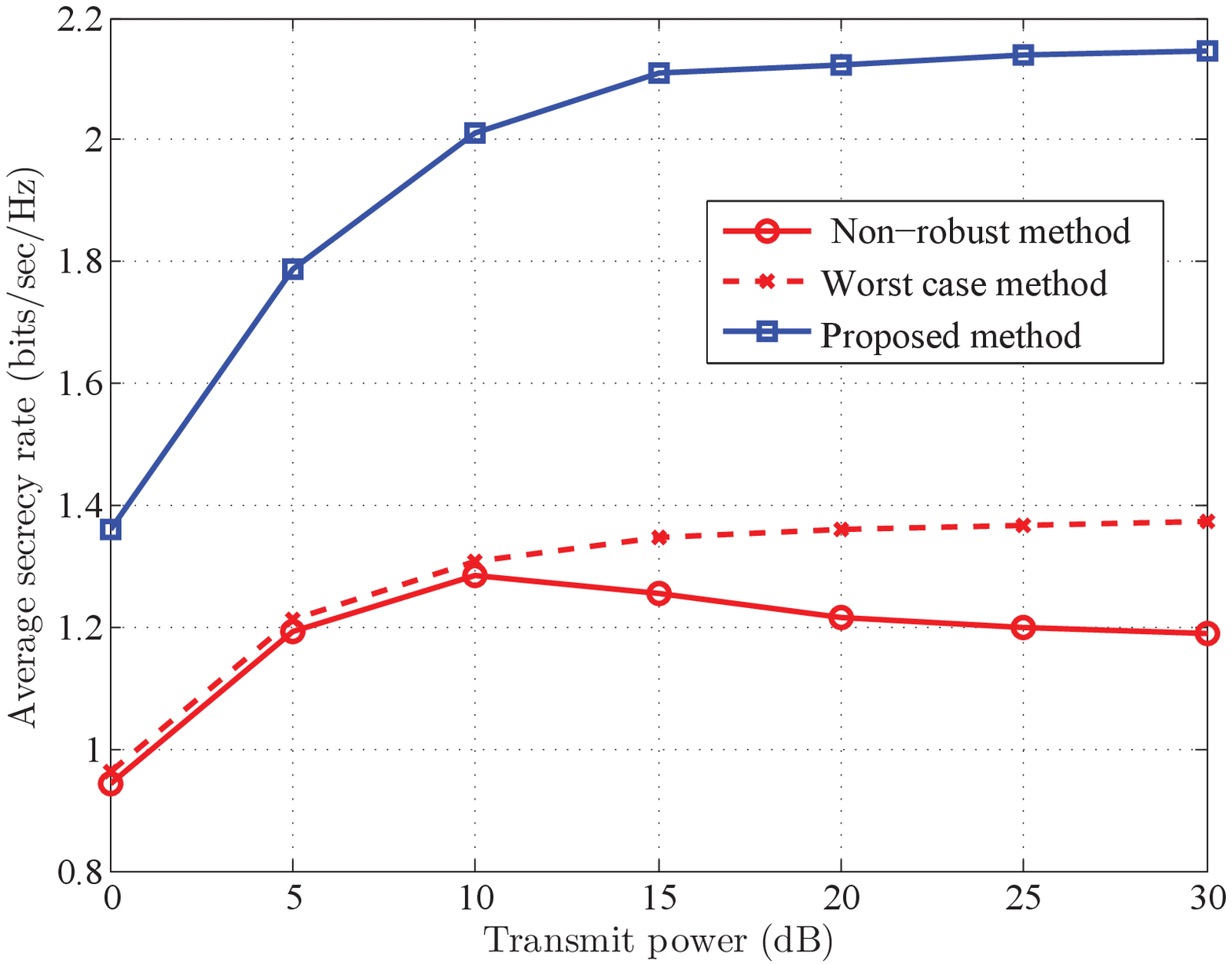}
      \vskip-0.2cm\centering {\footnotesize (b)}
    \end{minipage}\hfill
    \begin{minipage}[b]{0.45\textwidth}
      \centering
      \includegraphics[height=5cm,width=7cm]{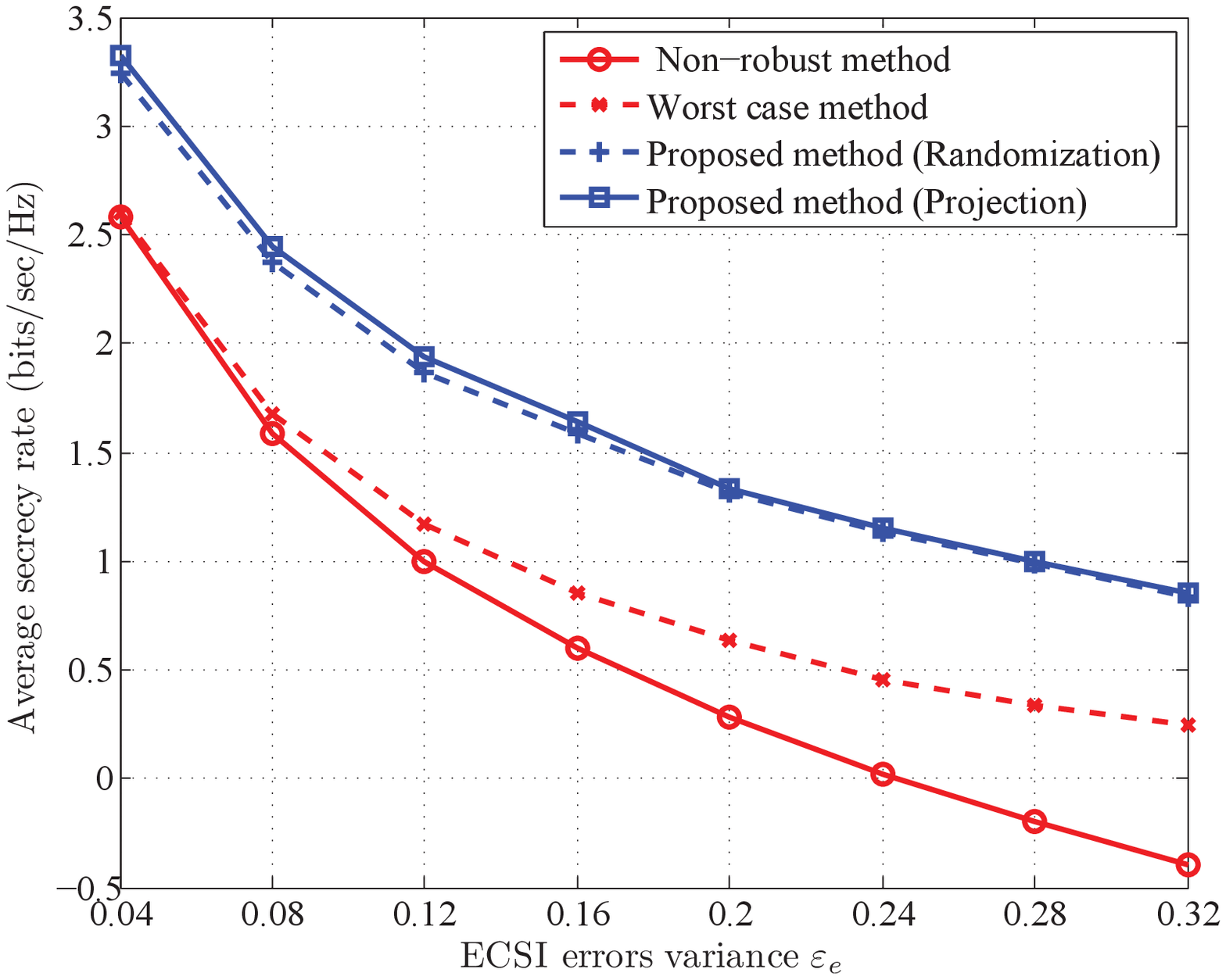}
      \vskip-0.2cm\centering {\footnotesize (c)}
    \end{minipage}
 \caption{(a)~The empirical CDF of  secrecy rate with  $R=3$ (bits/sec/Hz), ${\varepsilon _e} = 0.2$ and ${\overline p _{out}} = 0.05$; (b)~Average secrecy rate  versus transmit power with ${\varepsilon _e} = 0.1$; (c)~Average secrecy rate  versus ECSI errors variance $\varepsilon_e$.}
  \label{fig: Scenario_2} 
\end{figure}

The empirical  CDF  of the achieved secrecy rate for the problem \eqref{eq:scenario2} are plotted in Fig. \ref{fig: Scenario_2} (a). We set the target rate as  $R=3$ (bits/sec/Hz), set the ECSI error variance as ${\varepsilon _e} = 0.2$, and set the outage probability as ${\overline p _{out}} = 0.05$. Clearly, the non-robust design cannot satisfy the outage constraint, and about  $60\%$  of the rates are below the target rate $R=3$ (bits/sec/Hz). On the other hand, the achieved secrecy rates of both the  worst case method and the  proposed method  satisfy the outage constraint, but the proposed method is less conservative and achieves a better overall performance.  Fig. \ref{fig: Scenario_2} (b)  plots the secrecy rates of the
various methods against the transmit power with  ECSI error variance   ${\varepsilon _e} = 0.1$. Once again, the
secrecy rate performance of the proposed method is better than those of the other methods.
Moreover, we observe that for non-robust method, the rate
is not monotonically increasing with respect to the transmit power. This is because when the design does not take channel uncertainties into consideration, increasing the power may also help improve eavesdroppers' receptions \cite{Li3}.
Fig. \ref{fig: Scenario_2} (c) shows the average  secrecy  rate   versus
ECSI error variance $\varepsilon_e$.
For the proposed method, we further compare the Gaussian Randomization Procedure (Proposed method (Randomization)) with the Projection Approximation Procedure (Proposed method (Projection)).
It can be  observed that  larger  CSI error variance results in lower rate, and
that the proposed method has much higher rate than the worst case design method  and non-robust method over the whole CSI errors variance range. Moreover,  the performance of the Projection
Approximation Procedure is slightly better than that of the Gaussian Randomization Procedure, especially when the CSI error variance is small.

%


%
%
%
%
%
%

%

\subsection{Simulation Results for Scenario 3}

 We again assume that all channels are in Rayleigh flat fading, i.e., ${\widehat
{\mathbf{{h}}}} \sim \mathcal{CN}\left(
{\mathbf{0},\mathbf{I}} \right)$, and
${\widehat{\mathbf{g}}_k} \sim \mathcal{C}\mathcal{N}\left( {{\mathbf{0}},{\mathbf{I}}} \right)$, $\forall k \in \mathcal{K} $. The variances of LCSI and ECSI  errors   are
${\mathbf{E}_{b}}= \varepsilon_b   \times
{\mathbf{I}}$,    ${{\mathbf{E}}_{e,k}} = {\varepsilon _e} \times {\mathbf{I}}$, $\forall k \in \mathcal{K} $, respectively, where the parameters  $\varepsilon _b  \geq  0$ and $\varepsilon _e  \geq  0$   represent CSI error variances. Similarly as in the previous subsection, we use the worst case design method and the non-robust method developed in \cite{Li3} for comparison.

\begin{figure}[h]
        \begin{minipage}[b]{0.45\textwidth}
      \centering
      \includegraphics[height=5cm,width=7cm]{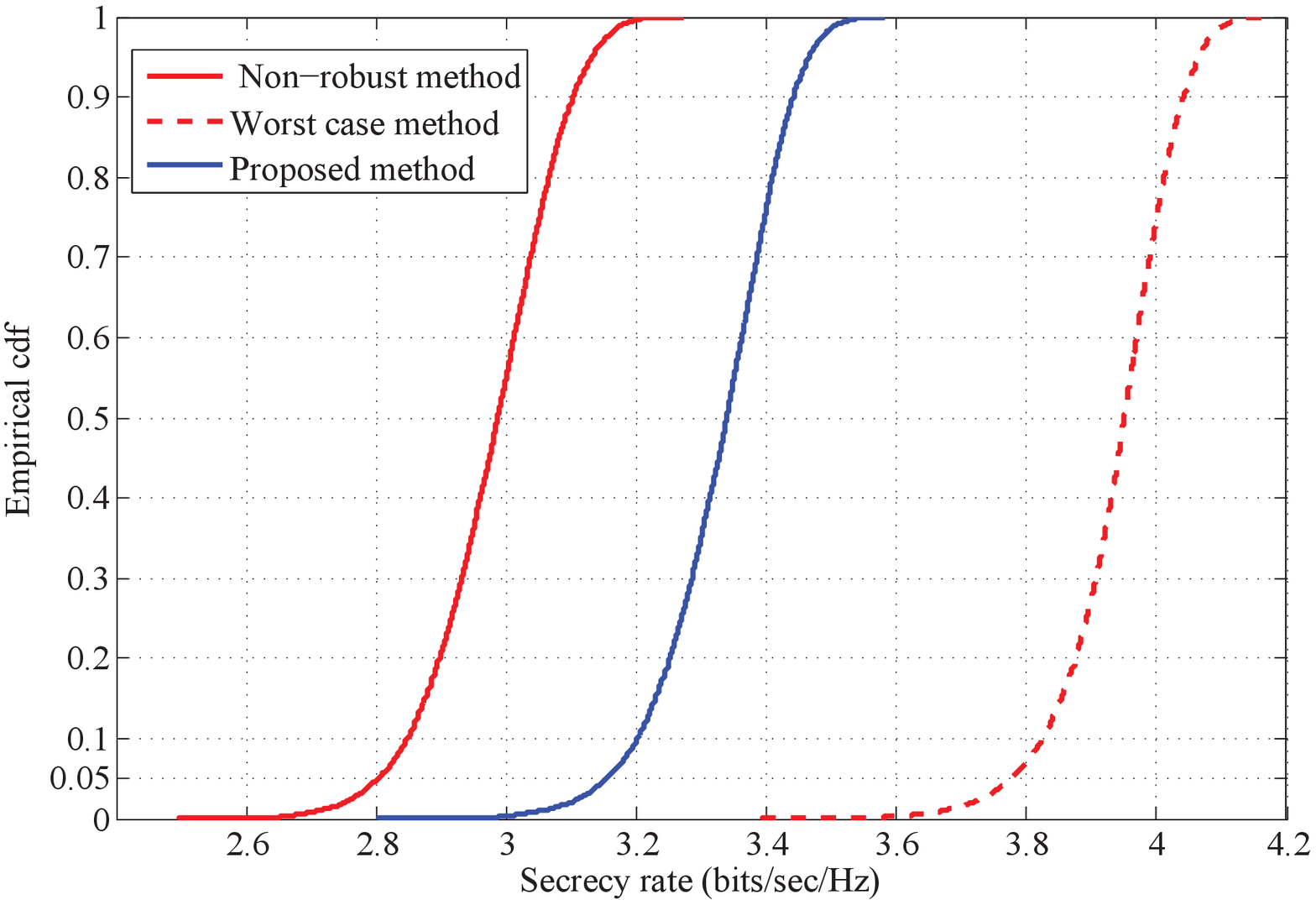}
      \vskip-0.2cm\centering {\footnotesize (a)}
    \end{minipage}
            \begin{minipage}[b]{0.45\textwidth}
      \centering
      \includegraphics[height=5cm,width=7cm]{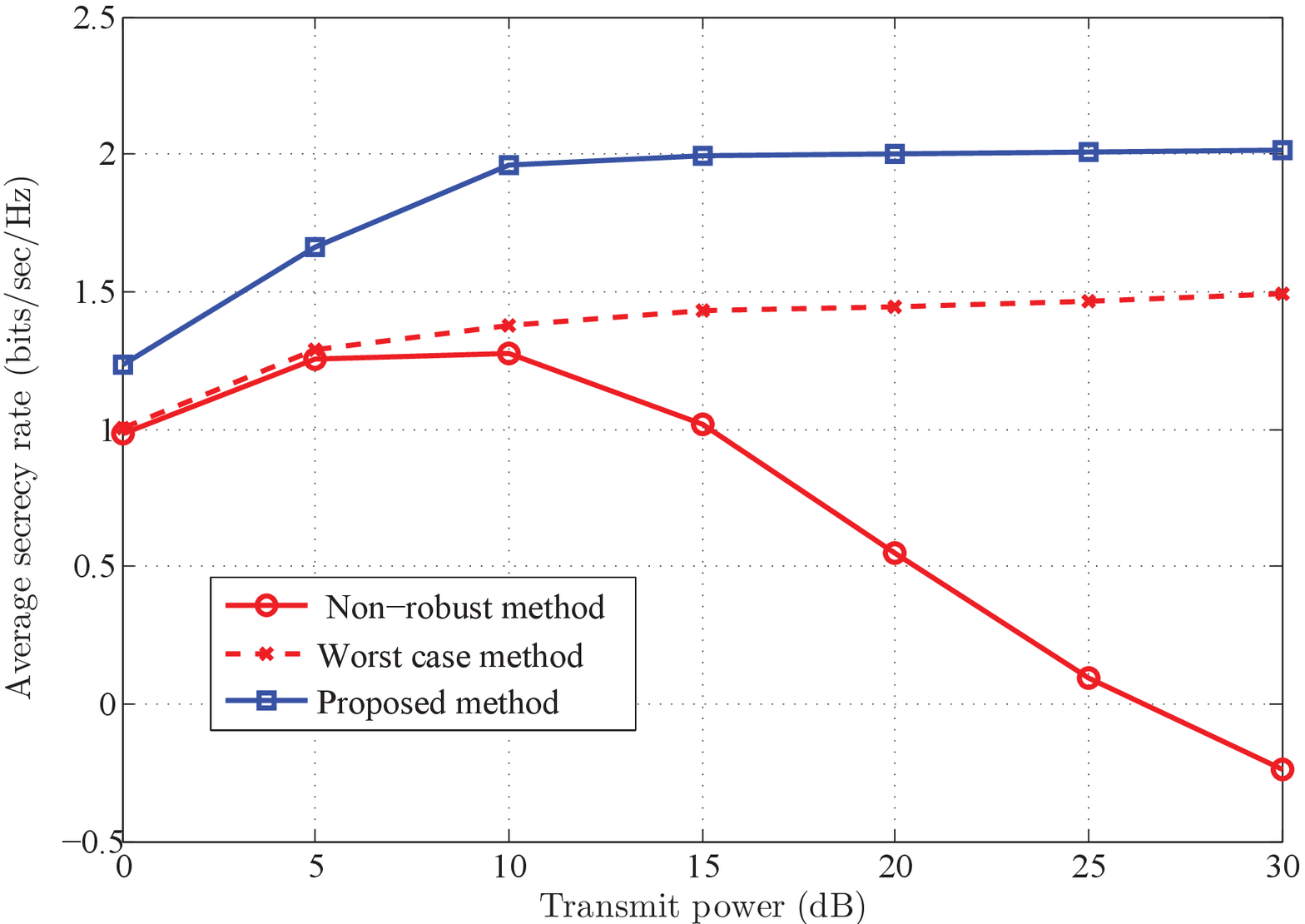}
      \vskip-0.2cm\centering {\footnotesize (b)}
    \end{minipage}
    \begin{minipage}[b]{0.45\textwidth}
      \centering
      \includegraphics[height=5cm,width=7cm]{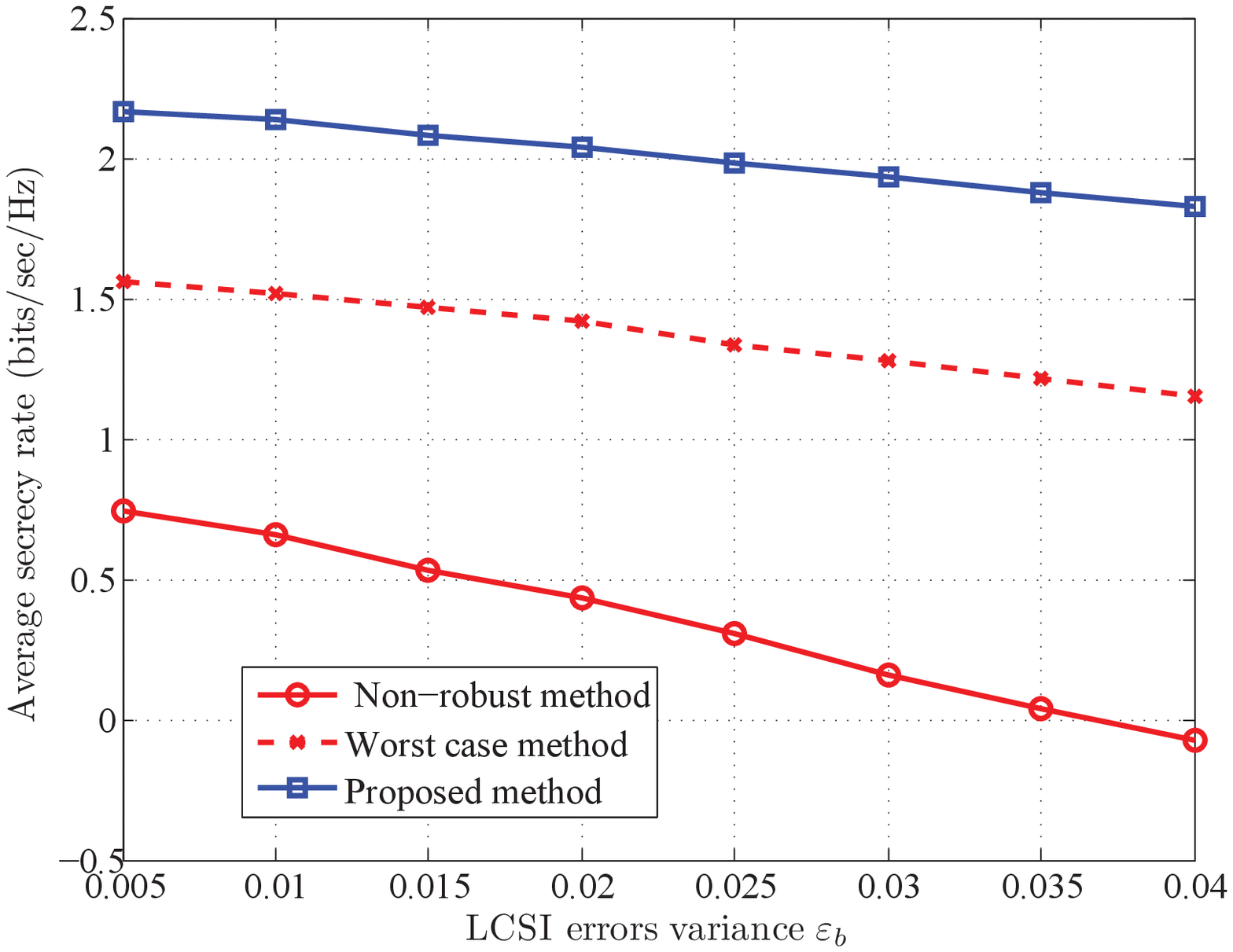}
      \vskip-0.2cm\centering {\footnotesize (c)}
    \end{minipage}\hfill
    \begin{minipage}[b]{0.45\textwidth}
      \centering
      \includegraphics[height=5cm,width=7cm]{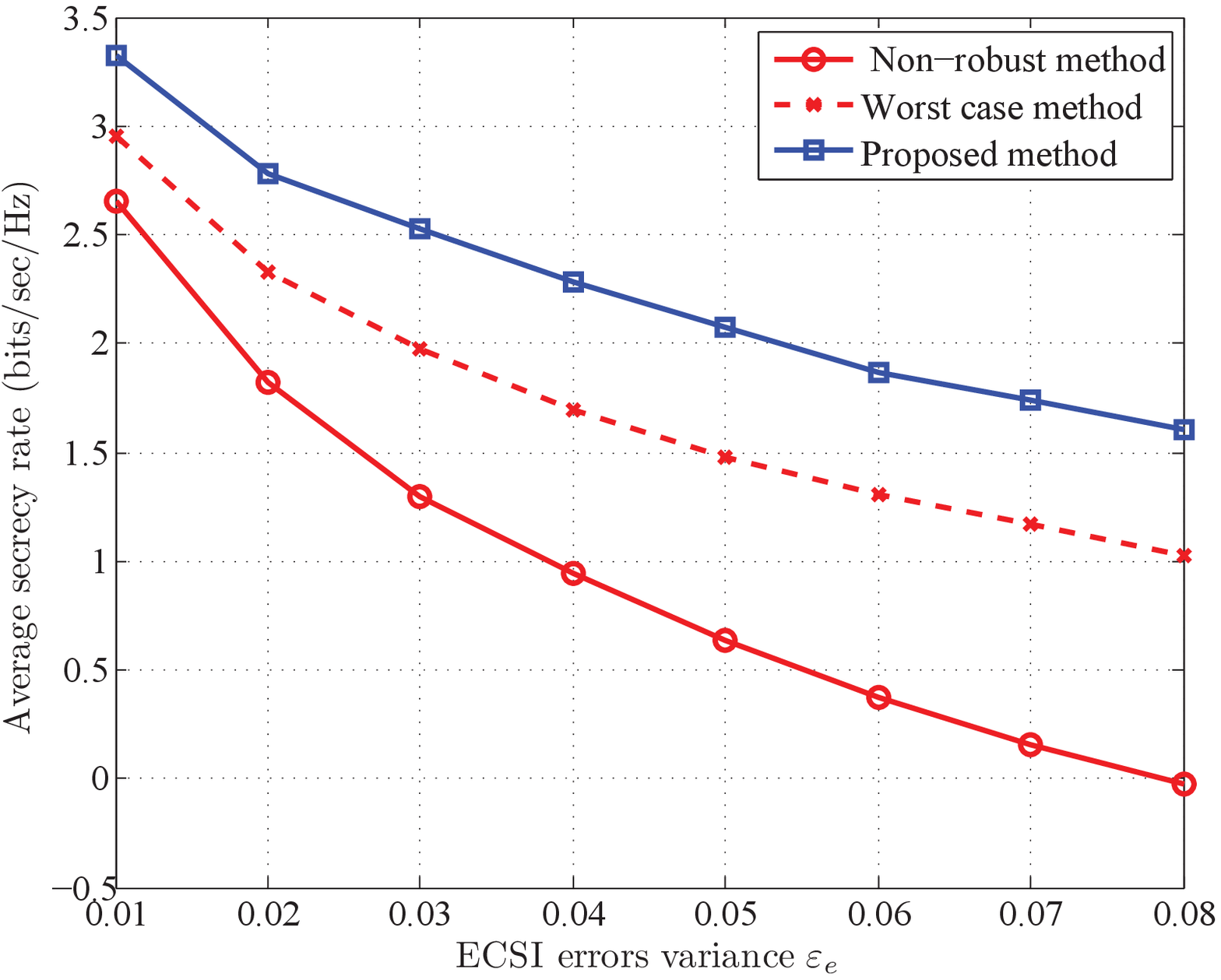}
      \vskip-0.2cm\centering {\footnotesize (d)}
    \end{minipage}
 \caption{(a)~The empirical CDF of  secrecy rate with $R=3$ (bits/sec/Hz), ${\varepsilon _b} = 0.005$, ${\varepsilon _e} = 0.2$ and ${\overline p _{out}} = 0.05$;(b)~Average secrecy rate  versus transmit power with ${\varepsilon _b} = 0.01$ and  ${\varepsilon _e} = 0.05$; (c)~Average secrecy rate  versus LCSI errors variance $\varepsilon_b$ with ${\varepsilon _e} = 0.05$; (d)~Average secrecy rate  versus ECSI errors variance $\varepsilon_e$  with ${\varepsilon _b} = 0.01$.}
  \label{fig: Scenario_3} 
\end{figure}

 The empirical  CDF   of the achieved secrecy rate for different algorithms are plotted in Fig. \ref{fig: Scenario_3} (a), where the target rate, the  LCSI errors variance  and the ECSI error variance is given by $R=3$ (bits/sec/Hz), ${\varepsilon _b} = 0.005$ and  ${\varepsilon _e} = 0.2$, respectively. As can be observed from the figure, the achieved secrecy rates of both the  worst case method and the proposed method satisfy the outage constraint (${\overline p _{out}} = 0.05)$, while the proposed method is less conservative than the worst case method. On the other hand, the non-robust design cannot satisfy the outage constraint, where  about  $55\%$  of the resulting secrecy rates fall below the  target rate $R=3$ (bits/sec/Hz).
Fig. \ref{fig: Scenario_3} (b)  plots the secrecy rates of various methods against the transmit power with  LCSI error variance  ${\varepsilon _b} = 0.01$ and  ECSI error variance  ${\varepsilon _e} = 0.05$. Not surprisingly, the
secrecy rate performance of the proposed method is better than those of the other methods. Moreover, we observe that the rate achieved by the non-robust method increases at first and then drops sharply, a phenomenon that has also been observed in Fig. \ref{fig: Scenario_2} (b).
Fig. \ref{fig: Scenario_3} (c)  (resp. Fig. \ref{fig: Scenario_3} (d)) presents the results of average secrecy rates of the various methods versus
LCSI error variance $\varepsilon_b$ (resp. ECSI error variance $\varepsilon_e$), with fixed ECSI error variance $\varepsilon_e=0.05$ (resp. LCSI error variance $\varepsilon_b=0.01$).   As can be seen from both figures, the secrecy rates of all  the three methods decease as the channel error variance increases, and the proposed method yields the best average secrecy rate.



 \section{Conclusions}

In this work, we focus on the design of robust secrecy beamforming strategies
for MISO wiretap channel.  We first formulate the general design problem as outage probability constrained optimization problem, and then develop different algorithms for computing high-quality solutions under various assumptions of CSI uncertainties. We show that when statistical ECSI and perfect LCSI are available, the chance constrained program can be solved to global optimality. For other two scenarios of CSI uncertainty, we propose to use a relaxation-restriction approach that can effectively obtain high-quality solutions for the difficult chance constrained program. Simulation results are provided to demonstrate the superior
performance of the proposed methods, both in terms of robustness and achievable secrecy rate. 


%
%
\appendices
 \section{Proof of Proposition 2}


 We note that there are $K$ constraints, and they share the same ${\mathbf{h}}$. In order to obtain the sufficient condition,   we first consider ${\mathbf{w}} = {\mathbf{h}}$ and we have
\begin{align}{{\mathbf{h}}^H}\left( {{{\mathbf{G}}_k}{\text{ln}}{p_{k,out}} + \frac{{\delta _{e,k}^2}}
{{\delta _b^2{2^R}}}{\mathbf{h}}{{\mathbf{h}}^H}} \right){\mathbf{h}} \geq  \delta _{e,k}^2\left( {1 - \frac{1}
{{{2^R}}}} \right).\label{eq:47}
\end{align}
The inequality \eqref{eq:47} can be reformulated as
\begin{align}
\frac{{\delta _{e,k}^2}}
{{\delta _b^2{2^R}}}{\left\| {\mathbf{h}} \right\|^4} \geq   - {{\mathbf{h}}^H}{{\mathbf{G}}_k}{\mathbf{h}}{\text{ln}}{p_{k,out}} - \delta _{e,k}^2\left( {1 - \frac{1}
{{{2^R}}}} \right).\label{eq:48}
\end{align}
Let $\rho \left( {{{\mathbf{G}}_k}} \right)$ denote the largest eigenvalue of the matrix ${{\mathbf{G}}_k}$, then we have
\begin{align}
\frac{{\delta _{e,k}^2}}
{{\delta _b^2{2^R}}}{\left\| {\mathbf{h}} \right\|^4} &\geq   - \rho \left( {{{\mathbf{G}}_k}} \right){\left\| {\mathbf{h}} \right\|^2}{\text{ln}}{p_{k,out}} - \delta _{e,k}^2\left( {1 - \frac{1}
{{{2^R}}}} \right).\label{eq:49}
\end{align}
This establishes \eqref{eq:47}.


Next, we show the necessary condition. If each of the largest eigenvalue of the  matrices $\left\{ {{{\mathbf{G}}_{k}}\ln {p_{k,out}} + \frac{{\delta _{e,k}^2}}
{{\delta _b^2{2^R}}}{\mathbf{h}}{{\mathbf{h}}^H}} \right\}$, $\forall k \in \mathcal{K}$  is  non-positive,  we have
\begin{align}{{\mathbf{G}}_k}{\text{ln}}{p_{k,out}} + \frac{{\delta _{e,k}^2}}
{{\delta _b^2{2^R}}}{\mathbf{h}}{{\mathbf{h}}^H} \preceq {\mathbf{0}},\forall \,k \in \mathcal{K}.\label{eq:52}
\end{align}
As a result, the left part of constraint  \eqref{eq:15b} is non-positive:
\begin{align}{{\mathbf{w}}^H}\left( {{{\mathbf{G}}_k}{\text{ln}}{p_{k,out}} + \frac{{\delta _{e,k}^2}}
{{\delta _b^2{2^R}}}{\mathbf{h}}{{\mathbf{h}}^H}} \right){\mathbf{w}} \leq  0,~\forall ~{\mathbf{w}}.\label{eq:53}
\end{align}
For $R > 0$, the right part  of constraint  \eqref{eq:15b} is positive: $\delta _{e,k}^2\left( {1 - \frac{1}
{{{2^R}}}} \right) > 0.$

Hence the constraint  \eqref{eq:15b}  cannot hold, thus the contradiction is established.

Therefore,  if problem \eqref{eq:power_min} under scenario 1 is feasible, then the largest eigenvalues of each of the  matrices $\left\{ {{{\mathbf{G}}_{k}}\ln {p_{k,out}} + \frac{{\delta _{e,k}^2}}{{\delta _b^2{2^R}}}{\mathbf{h}}{{\mathbf{h}}^H}} \right\}$, $\forall\; k \in \mathcal{K}$  must be  positive.

\section{Proof of Theorem 1}
The Lagrangian function for problem \eqref{eq:scenario1_SDR} is given by
 \begin{align}
&L\left( {\mathbf{W}} \right) = {\text{Tr}}\left( {\mathbf{W}} \right) - {\text{Tr}}\left( {{\mathbf{XW}}} \right) + \sum\limits_{k = 1}^K {{x_k}\left( {\delta _{e,k}^2 - \frac{{\delta _{e,k}^2}}
{{{2^R}}} - {\text{Tr}}\left( {\left( {{{\mathbf{G}}_k}\ln {p_{k,out}} + \frac{{\delta _{e,k}^2}}
{{\delta _b^2{2^R}}}{\mathbf{h}}{{\mathbf{h}}^H}} \right){\mathbf{W}}} \right)} \right)},  \label{eq:55}
\end{align}
where ${\mathbf{X}} \in {\mathbb C}^{{N_t}}$
  is the Lagrangian  dual variable for the constraint ${\mathbf{X}} \succeq {\mathbf{0}}$, and ${{x_k}}$, $\forall k \in \mathcal{K}$
are  the Lagrangian dual variables for the constraint  \eqref{eq:15b}.

The corresponding KKT conditions are shown to be
\begin{subequations}
\begin{align}
  &{\mathbf{I}} - {\mathbf{X}} - \sum\limits_{k = 1}^K {{x_k}\left( {{{\mathbf{G}}_k}\ln {p_{k,out}} + \frac{{\delta _{e,k}^2}}
{{\delta _b^2{2^R}}}{\mathbf{h}}{{\mathbf{h}}^H}} \right)}  = {\mathbf{0}}, \label{eq:56a} \\
  &{\text{Tr}}\left( {\left( {{{\mathbf{C}}_{e,k}}\ln {p_{k,out}} + \frac{{\delta _{e,k}^2}}
{{\delta _b^2{2^R}}}{\mathbf{h}}{{\mathbf{h}}^H}} \right){\mathbf{W}}} \right) \geq  \delta _{e,k}^2\left( {1 - \frac{1}
{{{2^R}}}} \right),\forall k \in \mathcal{K}, \label{eq:56b} \\
  &{\mathbf{XW}} = {\mathbf{0}}, \label{eq:56c} \\
  &{\mathbf{W}} \succeq {\mathbf{0}},{\mathbf{X}} \succeq {\mathbf{0}},{x_k} \geq  0,\forall k \in \mathcal{K}. \label{eq:56d}
\end{align}
\end{subequations}

Note that in general, \eqref{eq:scenario1_SDR} satisfies Slater's constraint qualification condition: If \eqref{eq:scenario1_SDR} has a feasible point, then one can prove,
by construction, that there exists a strictly feasible point for \eqref{eq:scenario1_SDR}.
As a result, strong duality holds and the KKT conditions are the
necessary conditions for a primal-dual point ${\mathbf{W}}{\text{,}}{\mathbf{X}}{\text{,}}\left\{ {{x_i}} \right\}$
  to be
optimal.

We rewrite \eqref{eq:56a} as
\begin{align}{\mathbf{X}} = {\mathbf{I}} - \sum\limits_{k = 1}^K {{x_k}{{\mathbf{G}}_k}\ln {p_{k,out}}}  - \left( {\sum\limits_{k = 1}^K {{x_k}\frac{{\delta _{e,k}^2}}
{{\delta _b^2{2^R}}}} } \right){\mathbf{h}}{{\mathbf{h}}^H}.\label{eq:57}
\end{align}

Since ${\mathbf{I}} - \sum\limits_{k = 1}^K {{x_k}{{\mathbf{G}}_k}\ln {p_{k,out}}}  \succ {\mathbf{0}},$
 and
${\text{rank}}\left( {\sum\limits_{k = 1}^K {{x_k}\frac{{\delta _{e,k}^2}}
{{\delta _b^2{2^R}}}} {\mathbf{h}}{{\mathbf{h}}^H}} \right) = 1,$
 we have
${\text{rank}}\left( {\mathbf{X}} \right) \geq  {N_t} - 1.$

Since ${\mathbf{XW}} = {\mathbf{0}}$, we have  ${\text{rank}}\left( {\mathbf{W}} \right) \leq 1$.
If ${\text{rank}}\left( {\mathbf{W}} \right) = 0$, then ${\mathbf{W}} = {\mathbf{0}}$. However, the constraint \eqref{eq:15b} violates when
$R > 0$.
Hence, ${\text{rank}}\left( {\mathbf{W}} \right) =1$.

\section{Proof of Proposition 3}
To simplify notation, we assume that ${\mathbf{W}}$ is the optimal solution of the problem \eqref{eq:scenario2_SDP} in this proof.
Let ${\mathbf{P}}$ denote the projection matrix of vector ${{\mathbf{W}}^{{1 \mathord{\left/
 {\vphantom {1 2}} \right.
 \kern-\nulldelimiterspace} 2}}}{\mathbf{h}}$:
\begin{align}{\mathbf{P}} = \frac{{{{\mathbf{W}}^{{1 \mathord{\left/
 {\vphantom {1 2}} \right.
 \kern-\nulldelimiterspace} 2}}}{\mathbf{h}}{{\left( {{{\mathbf{W}}^{{1 \mathord{\left/
 {\vphantom {1 2}} \right.
 \kern-\nulldelimiterspace} 2}}}{\mathbf{h}}} \right)}^H}}}
{{{{\left\| {{\mathbf{h}}{{\mathbf{W}}^{{1 \mathord{\left/
 {\vphantom {1 2}} \right.
 \kern-\nulldelimiterspace} 2}}}} \right\|}^2}}} = \frac{{{{\mathbf{W}}^{{1 \mathord{\left/
 {\vphantom {1 2}} \right.
 \kern-\nulldelimiterspace} 2}}}{{\mathbf{h}}^H}{\mathbf{h}}{{\mathbf{W}}^{{1 \mathord{\left/
 {\vphantom {1 2}} \right.
 \kern-\nulldelimiterspace} 2}}}}}
{{{{\left( {{{\mathbf{W}}^{{1 \mathord{\left/
 {\vphantom {1 2}} \right.
 \kern-\nulldelimiterspace} 2}}}{\mathbf{h}}} \right)}^H}{{\mathbf{W}}^{{1 \mathord{\left/
 {\vphantom {1 2}} \right.
 \kern-\nulldelimiterspace} 2}}}{\mathbf{h}}}}.\label{eq:61}\end{align}

We construct a new rank one   solution  $\widehat{\mathbf{W}}$ as
$\widehat{\mathbf{W}} = {{\mathbf{W}}^{{1 \mathord{\left/
 {\vphantom {1 2}} \right.
 \kern-\nulldelimiterspace} 2}}}{\mathbf{P}}{{\mathbf{W}}^{{1 \mathord{\left/
 {\vphantom {1 2}} \right.
 \kern-\nulldelimiterspace} 2}}}.$
Firstly, it is easy to see that the new solution ${\widehat{\mathbf{W}}}$ is a rank one matrix.
Then let us check the value of the objective function,
${\mathbf{W}} - \widehat{\mathbf{W}}{\text{ = }}{{\mathbf{W}}^{{1 \mathord{\left/
 {\vphantom {1 2}} \right.
 \kern-\nulldelimiterspace} 2}}}\left( {{\text{I}} - {\mathbf{P}}} \right){{\mathbf{W}}^{{1 \mathord{\left/
 {\vphantom {1 2}} \right.
 \kern-\nulldelimiterspace} 2}}} \succeq \mathbf{0}.$
Thus ${\text{Tr}}\left( {\widehat{\mathbf{W}}} \right) \leq  {\text{Tr}}\left( {\mathbf{W}} \right)$,  which means   the value of the objective function will not increase. Finally,  let us check whether the constraint \eqref{eq:15b} is    satisfied for the new solution $\widehat{\mathbf{W}}$.
The constraint \eqref{eq:15b} can be equivalently reformulated as
\begin{align}
  &\Pr \Bigg{\{} {\underbrace {{{\log }_2}\left( {1 + \frac{{{{\mathbf{h}}^H}{\mathbf{Wh}}}}
{{\delta _b^2}}} \right)}_{{\text{Part 1}}} - \underbrace {{{\log }_2}\left( {1 + \frac{{{\mathbf{g}}_k^H{\mathbf{W}}{{\mathbf{g}}_k}}}
{{\delta _{e,k}^2}}} \right)}_{{\text{Part 2}}}}{ \geq  R} \Bigg{\}}\geq  1 - {p_{k,out}}. \label{eq:64}
\end{align}

Substituting $\widehat{\mathbf{W}}$ into the Part 1, we have
\begin{align}
  &{{\mathbf{h}}^H}\widehat{\mathbf{W}}{\mathbf{h}}
   ={{\mathbf{h}}^H}{{\mathbf{W}}^{{1 \mathord{\left/
 {\vphantom {1 2}} \right.
 \kern-\nulldelimiterspace} 2}}}{\mathbf{P}}{{\mathbf{W}}^{{1 \mathord{\left/
 {\vphantom {1 2}} \right.
 \kern-\nulldelimiterspace} 2}}}{\mathbf{h}}
   = \frac{{{{\mathbf{h}}^H}{{\mathbf{W}}^{{1 \mathord{\left/
 {\vphantom {1 2}} \right.
 \kern-\nulldelimiterspace} 2}}}{{\mathbf{W}}^{{1 \mathord{\left/
 {\vphantom {1 2}} \right.
 \kern-\nulldelimiterspace} 2}}}{{\mathbf{h}}^H}{\mathbf{h}}{{\mathbf{W}}^{{1 \mathord{\left/
 {\vphantom {1 2}} \right.
 \kern-\nulldelimiterspace} 2}}}{{\mathbf{W}}^{{1 \mathord{\left/
 {\vphantom {1 2}} \right.
 \kern-\nulldelimiterspace} 2}}}{\mathbf{h}}}}
{{{{\left( {{{\mathbf{W}}^{{1 \mathord{\left/
 {\vphantom {1 2}} \right.
 \kern-\nulldelimiterspace} 2}}}{\mathbf{h}}} \right)}^H}{{\mathbf{W}}^{{1 \mathord{\left/
 {\vphantom {1 2}} \right.
 \kern-\nulldelimiterspace} 2}}}{\mathbf{h}}}}
   = {{\mathbf{h}}^H}{\mathbf{Wh}} \label{eq:65d}.
\end{align}
Hence, the value of the Part 1  remains the same ${\mathbf{W}}$ is replaced with $\widehat{\mathbf{W}}$.
Moreover,  we have
\begin{align}
  &{\mathbf{g}}_k^H{\mathbf{W}}{{\mathbf{g}}_k} - {\mathbf{g}}_k^H\widehat{\mathbf{W}}{{\mathbf{g}}_k}
   = {\mathbf{g}}_k^H\left( {{\mathbf{W}} - {{\mathbf{W}}^{{1 \mathord{\left/
 {\vphantom {1 2}} \right.
 \kern-\nulldelimiterspace} 2}}}{\mathbf{P}}{{\mathbf{W}}^{{1 \mathord{\left/
 {\vphantom {1 2}} \right.
 \kern-\nulldelimiterspace} 2}}}} \right){{\mathbf{g}}_k}
   = {\mathbf{g}}_k^H\left( {{{\mathbf{W}}^{{1 \mathord{\left/
 {\vphantom {1 2}} \right.
 \kern-\nulldelimiterspace} 2}}}\left( {{\text{I}} - {\mathbf{P}}} \right){{\mathbf{W}}^{{1 \mathord{\left/
 {\vphantom {1 2}} \right.
 \kern-\nulldelimiterspace} 2}}}} \right){{\mathbf{g}}_k}
   \geq 0 \label{eq:66e}
\end{align}

Thus the value of the Part 2 will not increase if we replace ${\mathbf{W}}$ with $\widehat{\mathbf{W}}$.

Therefore,  the constraint \eqref{eq:15b} is still satisfied for the new rank one solution $\widehat{\mathbf{W}}$.

\end{document}